\documentclass[useAMS,usenatbib,aas_macros]{mn2e}
\usepackage{aas_macros}
\newcommand{\equ}[1]{eq.~(\ref{eq:#1})}
\newcommand{\Equ}[1]{Eq.~(\ref{eq:#1})}
\newcommand{\equnp}[1]{eq.~\ref{eq:#1}}
\newcommand{\se}[1]{\S\ref{sec:#1}}
\newcommand{\fig}[1]{Fig.~\ref{fig:#1}}
\newcommand{\Fig}[1]{Figure~\ref{fig:#1}}
\newcommand{\be}{\begin{equation}}
\newcommand{\ee}{\end{equation}}

\def\no{\noindent}

\newcommand{\msun}{M_\odot}
\newcommand{\lsun}{L_\odot}
\newcommand{\ifm}[1]{\relax\ifmmode#1\else$\mathsurround=0pt #1$\fi}
\newcommand{\kms}{\ifmmode\,{\rm km}\,{\rm s}^{-1}\else km$\,$s$^{-1}$\fi}
\newcommand{\hmpc}{\,\ifm{h^{-1}}{\rm Mpc}}

\newcommand{\kpc}{\,{\rm kpc}}
\newcommand{\Gyr}{\,{\rm Gyr}}

\newcommand{\ltsima}{$\; \buildrel < \over \sim \;$}
\newcommand{\lsim}{\lower.5ex\hbox{\ltsima}}
\newcommand{\gtsima}{$\; \buildrel > \over \sim \;$}
\newcommand{\gsim}{\lower.5ex\hbox{\gtsima}}
\newcommand{\prop}{\propto}

\newcommand{\gamef}{\gamma_{\rm eff}}
\newcommand{\gamc}{\gamma_{\rm crit}}
\newcommand{\lya}{Ly-$\alpha$\ }

\newcommand{\Ms}{M_{\rm s}}
\newcommand{\ms}{M_{\rm s}}
\newcommand{\mus}{\mu_{\rm s}}
\newcommand{\msc}{M_{{\rm s,crit}}}
\newcommand{\mc}{M_{\rm crit}}
\newcommand{\vc}{V_{\rm crit}}
\newcommand{\msh}{M_{\rm shock}}
\newcommand{\mst}{M_{\rm stream}}
\newcommand{\mfb}{M_{\rm fdbk}}
\newcommand{\mps}{M_{*}} 

\newcommand{\zc}{z_{\rm crit}}

\newcommand{\half}{\frac{1}{2}}

\newcommand{\omm}{\Omega_{\rm m}}

\newcommand{\oml}{\Omega_{\Lambda}}

\newcommand{\Vv}{V_{\rm v}}
\newcommand{\Mv}{M_{\rm v}}
\newcommand{\Rv}{R_{\rm v}}
\newcommand{\Tv}{T_{\rm v}}

\newcommand{\fb}{f_{\rm b}}

\newcommand{\rhou}{\rho_{\rm u}}

\newcommand{\tc}{t_{\rm cool}}
\newcommand{\tp}{t_{\rm comp}}
\newcommand{\tu}{t_{\rm univ}}

\newcommand{\eps}{\epsilon}
\newcommand{\rs}{r_{\rm s}}
\newcommand{\us}{u_{\rm s}}
\newcommand{\ust}{\tilde{u}_{\rm s}}
\newcommand{\fr}{f_r}
\newcommand{\fu}{f_u}

\title[Cold flows and galaxy bimodality]
{Galaxy Bimodality due to Cold Flows and Shock Heating}

\author[A. Dekel \& Y. Birnboim]
{Avishai Dekel \& Yuval Birnboim\\
Racah Institute of Physics, The Hebrew University, Jerusalem Israel\\
dekel@phys.huji.ac.il; yuval@phys.huji.ac.il}

\begin{document}

\pagerange{\pageref{firstpage}--\pageref{lastpage}} \pubyear{2002}

\maketitle

\label{firstpage}

%%%%%%%%%%%%%%%%%%%%%%%%%%%%%%%%%%%%%%
\begin{abstract}
We address the origin of the robust bi-modality observed in galaxy properties
about a characteristic stellar mass $\sim\!3\!\times 10^{10}\msun$.  
Less massive galaxies tend to be ungrouped {\it blue} star-forming discs, 
while more massive galaxies are typically grouped
{\it red} old-star spheroids. 
Color-magnitude data show a gap between the red and blue sequences, 
extremely red luminous galaxies already at $z\!\sim\!1$, 
a truncation of today's blue sequence above $L_*$, 
and massive starbursts at $z \sim 2-4$.
We propose that these features are driven by the {\it thermal\,} properties of 
the inflowing gas and their interplay with the clustering
and feedback processes, all functions of the dark-matter halo mass
and associated with a similar characteristic scale.  
In haloes below a critical shock-heating mass 
$\msh\!\lsim\!10^{12}\msun$, discs are built by {\it cold streams}, not 
heated by a virial shock, yielding efficient early star formation. 
It is regulated by supernova feedback into a long sequence of 
bursts in blue galaxies constrained to a ``fundamental line".  
Cold streams penetrating 
through hot media in $M\geq\!\msh$ haloes preferentially 
at $z\!\geq\!2$ lead to massive starbursts in $L>\!L_*$ galaxies.
At $z\!<\!2$, in $M>\!\msh$ haloes hosting 
groups, the gas is heated by a virial shock, and being dilute it becomes 
vulnerable to feedback from energetic sources such as AGNs.  This shuts off 
gas supply and prevents further star formation, leading by passive evolution
to ``red-and-dead" massive spheroids starting at $z\!\sim\!1$. 
A minimum in feedback efficiency near $\msh$ explains
the observed minimum in $M/L$ and the qualitative features of the 
star-formation history.  The cold flows provide a hint for solving the
angular-momentum problem. 
When these processes are incorporated in simulations 
they recover the main bi-modality features and solve other open puzzles.
\end{abstract}


%%%%%%%%%%%%%%%%%%%%%%%%%%%%%%%
\section{Introduction}
\label{sec:intro}

% bi-modality summary
Observations reveal a robust bi-modality in the galaxy population,
being divided into two classes, the ``blue" and ``red" sequences,
at a characteristic stellar mass $\msc \simeq 3\times 10^{10}\msun$.
This corresponds to a dark-halo mass $\mc \lsim 10^{12}\msun$
and a virial velocity $\vc \simeq 120\kms$ today.
%In a nut shell,
Less massive galaxies tend to be blue, star-forming discs residing
in the ``field".
Their properties are correlated along a ``fundamental line" of decreasing
surface brightness, internal velocity and metallicity with decreasing 
luminosity.
% down to the smallest dwarf galaxies.  
Galaxies above $\mc$ are dominated by
spheroids of red, old stars, with high surface brightness and
metallicity independent of luminosity. They tend to reside in 
the high-density environments of groups and clusters
and they preferentially host Active Galactic Nuclei (AGN).
%The mean halo mass-to-light ratio has a minimum at the critical mass.

% very blue and very red
Current models of galaxy formation have difficulties in reproducing
this bi-modality and the broad color distribution observed. In particular,
the extremely red bright ellipticals which start showing up already at 
$z\sim 1$ are not predicted. 
They require efficient star formation at earlier epochs,
followed by an effective shut-down of star formation in massive galaxies.
The observations also reveal very blue galaxies in excess
of the predictions, indicating repeating starbursts 
over the lifetimes of galaxies.
Today's blue sequence is non-trivially truncated at the
bright end, while at $z \geq 2$ there are indications for very luminous
starbursts in big objects, both posing severe theoretical challenges.

%-------------------------------------
\subsection{The observed bi-modality}

The bi-modality or transition in galaxy properties
is observed in many different ways.
We list here the main relevant observed features\footnote{Quoting only
sample references, making no attempt to be complete.} which we address
in this paper (\se{bimo}).

\no (a) {\bf Luminosity functions}. 
Blue galaxies dominate the stellar mass function below $\msc$
while red galaxies take over above it 
\citep[][in SDSS and 2MASS]{baldry04,bell03_2mass}.
The transition occurs slightly below $L_*$,
the characteristic luminosity of the brightest disc galaxies, 
beyond which the luminosity function drops. 

\no (b) {\bf Color-magnitude}.
A color bi-modality shows robustly in color-magnitude diagrams,
where the galaxies are divided into a blue sequence and a red sequence
separated by a gap.  In SDSS \citep{blanton04a,baldry04}, 
the gap is at $u-r \sim 2$.
The color distribution is non-trivially {\it broad}, with the red tip
stretching beyond $u-r=2.5$ and the blue tail reaching well below $u-r=1.0$.

\no (c) {\bf Star-formation rate}.
The current star-formation rate (SFR), and the typical age of the stellar
population, show a robust bi-modality about $\msc$.
The less massive galaxies are dominated by young populations, while the
more massive galaxies are dominated by old stars
\citep[][in SDSS and 2dF]{kauf03_pop,lahav03},
in agreement with the color bi-modality.
A similar bi-modality is seen in the gas-to-stellar mass fraction,
which is high in the blue sequence and low in the red sequence,
steeply increasing with stellar mass below $\msc$, and only moderately
so above it \citep[][in SDSS+2MASS]{kannappan04}.

\no (d) {\bf Color-magnitude at $z\sim 1$}.
The color bi-modality is similar back to $z \sim 1.5$ 
\citep[][in COMBO17]{bell04_combo}. %DEEP2 ? 
Extremely red massive galaxies (EROs) exist at the bright tip of the
red sequence already at $z\sim 1$ \citep[e.g.][]{moustakas04_red}.
Very blue small galaxies indicating starbursts show in the blue sequence
\citep[e.g.][]{ferguson98,fioc99}.

\no (e) {\bf Massive starbursts at high $z$}.
Very luminous and massive dusty objects are detected at $z\sim 2-4$,
indicating an excessive activity of star formation in surprisingly big objects
\citep[][LBG and SCUBA sources]{shapley04,smail02,chapman03_scuba,chapman04}. 

% out?
\no (f) {\bf Star-formation history}.
The cosmological history of star-formation rate has a broad maximum near
$z\sim 1-2$, followed by a sharp drop from $z\sim 1$ to $z=0$
\citep[e.g.][]{madau96_sfr,dickinson03_sfr,
hartwick04_sfr,giavalisco04_sfr,heavens04_sfr}.
Still, about half the stars in today's spirals seem to have formed after
$z \sim 1$, e.g. in luminous infrared galaxies (LIRGs) near $\msc$
\citep{hammer04}.
Massive galaxies tend to form their stars earlier than smaller galaxies
\citep[``downsizing"][]{thomas05}.

\no (g) {\bf Bulge-to-disc ratio}.
The galaxy bulge-to-disc ratio shows a transition from disc dominance in the
blue sequence below $\msc$ to spheroid dominance in the red sequence
\citep[][in SDSS]{kauf03_pop,blanton04a}.

\no (h) {\bf Environment dependence}.
The distributions in color and SFR depend strongly on
the galaxy density in the $\sim 1 Mpc$ vicinity:
the blue and red sequence galaxies tend to populate low and high 
density environments respectively
\citep[][in SDSS]{hogg03,kauf04_env,blanton04a,balogh04_env,blanton04b}.
The color-environment correlation is stronger than the  
morphology-density relation \citet{dressler80}.  

\no (i) {\bf Halo mass and HOD}.
The environment density is correlated with the mass of the host dark-matter
(DM) halo, where haloes less massive than $\sim 10^{12}\msun$ typically 
host one dominant galaxy each while more massive haloes tend to host 
groups and clusters of luminous galaxies, as quantified by the 
Halo Occupation Distribution \citep[HOD,][in 2dF, SDSS and in simulations]
{yan03_hod_2df,zehavi04_hod, krav04_hod}.
The environment dependence thus implies that the galaxy properties are 
correlated with the host-halo mass,
with the bi-modality at $\mc \lsim 10^{12}\msun$ \citep{blanton04b}.

\no (j) {\bf Hot halo gas}.
Ellipticals of $L_{\rm B} > 10^{10.5}\lsun$ show
a significant excess of X-ray flux plausibly associated with
hot halo gas \citep{ciotti91,mathews03}. 
Inter-galactic X-ray radiation is detected predominantly in
groups where the brightest galaxy is an elliptical.
Group properties have a transition near 
$\sigma_{\rm v} \sim 140\kms$ \citep{helsdon03,osmond04}.
%The luminosity of the brightest group galaxy is strongly correlated with the
%group halo mass for $M<6\times 10^{11}\msun$, and is weakly correlated with 
%it above this scale (based on SDSS, R. Wechsler, private communication).

\no (k) {\bf Luminosity/mass functions}.
The stellar-mass function has a ``knee" near $\msc$,
where the shallow $dn/d\Ms \prop \Ms^{-1}$ on the faint side
turns into an exponential drop.  In contrast, 
the dark-halo mass function is predicted by the standard $\Lambda$CDM model
to be $dn/dM \prop M^{-1.8}$ everywhere 
below $\sim 10^{13}\msun$.  
A match at $\msc$ requires a baryonic fraction $\Ms/M \sim 0.05$,
indicating gas loss, and associating 
$\msc$ with $\mc \simeq 6\times 10^{11}\msun$.   
The halo mass-to-light function has a minimum near $\mc$,
varying as $M/L \prop M^{-2/3}$ and $\prop M^{+1/2}$ below and above it
respectively, thus implying increased suppression of star formation away from
the critical mass on both sides
\citep*[][in a B-mag sample and in SDSS, 2MASS, 2dF]{marinoni02,
bell03_sam,bell03_baryon,  %ref tex problem xxx
yang03}.
%D
%Matching the luminosity function of central galaxies in haloes
%to the $\Lambda$CDM subhalo mass function indicates a similar transition 
%near $\mc$
%\citep[][;Wechsler \& White, in preparation]{tasitsiomi04}.
%Accordingly, the luminosity of brightest cluster galaxies, above $\mc$,
%are growing very slowly with cluster halo mass \citep{lin04}.

% out?
\no (l) {\bf Fundamental line vs.~plane}.
A transition is detected in the galaxy structural scaling relations near 
$\msc$, e.g., the surface brightness changes from 
$\mus \prop \Ms^{0.6}$ at lower masses to $\mus \sim const.$ at the bright end
\citep[][in SDSS]{kauf03_pop}.  The correlation below $\msc$ is part of 
the ``fundamental line" relating stellar mass, radius and velocity
over five decades in $\Ms$ \citep[e.g.][]{dw03}.
%{\bf Metallicity}.
The mean metallicity shows a transition near a similar mass scale 
from $Z \sim \Ms^{0.4}$ to $Z \sim const.$
\citep[][in SDSS and the Local Group]{tremonti04,dw03}.

\no (m) {\bf AGN}.
Black hole masses are correlated with their host spheroid properties
\citep[e.g.][]{tremaine02}.
The optical AGN population, with high
accretion rate and SFR, peaks near $\msc$ with little AGN activity at smaller
masses, and is associated with black-hole masses $\lsim 10^8\msun$
\citep[][in SDSS]{kauf03_agn}. 
Radio-loud AGNs, uncorrelated with the optical activity and the SFR, 
dominate in larger haloes hosting $\sim 10^{8-9}\msun$ black holes
(G. Kauffmann, private communication).

%-------------------------------
\subsection{Key Physical Processes}

% processes
The bi-modality imprinted on almost every global
property of galaxies deserves a simple theoretical understanding.
We propose that the main source of the bi-modality is the transition
from cold flows to virial shock-heating at a critical scale, 
in concert with feedback processes and gravitational clustering
that emphasize the same characteristic scale. 
We address the cross-talk between these processes, 
and integrate them into a scenario which
attempts to address simultaneously the variety of observed phenomena.
The key processes are:

\no (a) {\bf Cold infall vs.~hot medium}.
The thermal behavior of the gas as it falls through the halo  
is qualitatively different below and above a critical mass scale
of $\msh \lsim 10^{12}\msun$ \citep{bd03,keres04}. 
In less massive haloes, the disc is built by cold flows ($\sim 10^{4-5}$K),
which are likely to generate early bursts of star formation. 
In more massive haloes, the infalling gas is first heated by stable
shocks to near the virial temperature ($\sim 10^6$K).
Near and above $\msh$ at $z \geq 2$ (and preferentially
in isolated galaxies), streams of dense cold gas penetrate 
through the dilute shock-heated medium \citep{fardal01,krav03,keres04}
(discussed in \se{shock_spheri}-\se{simu}). 

\no (b) {\bf Gravitational clustering}.
Non-linear gravitational clustering of the DM 
occurs on a characteristic mass scale, $\mps$,
marking the typical haloes forming at a given epoch and
the lower bound for groups of galaxies.
We point out that the clustering scale, which varies rapidly with cosmological
time, happens to 
coincide with $\msh$ at $z \sim 1$, and the interplay
between these scales plays a role in determining the galaxy
properties (\se{simu}).

\no (c) {\bf Feedback}.
We argue that the feedback processes affecting galaxy evolution are relatively
ineffective near $\msh$, largely due to the
shock-heating process itself, and they therefore help emphasizing the imprint
of this scale on the galaxy properties.
Supernova and other feedback processes 
regulate star formation in the blue sequence below $\msh$.
Feedback by AGNs, or other sources, 
becomes efficient in haloes more massive than $\msh$, because it
preferentially affects the dilute shock-heated medium
and may prevent it from ever cooling and forming stars.
effects (\se{feed}).  

% fundamental difference
We show how the introduction of shock-stability physics crystallizes
our understanding of the origin of the characteristic scales of galaxies.
We argue that the combination of shock heating, feedback and clustering
introduces a new feature in galaxy-formation modeling --- 
{\it a complete suppression of cold gas supply in haloes above a critical
mass after a critical redshift}. 
This could be the key to solving many of the open questions
posed by the observations,
focusing on the bright-end truncation of the luminosity
function, the appearance of very red bright galaxies already at $z\sim 1$
at the expense of big blue galaxies,
and the indications for massive starbursts at higher redshifts. 
We note that some of the issues 
have been addressed in parallel, in a similar spirit 
and in different ways, by
\citet{bd03}, \citet{benson03}, \citet{keres04} and \citet{binney04}.
We make here a more thorough investigation into the cold flows versus shock
heating phenomenon, relate it to the feedback and clustering processes,
and attempt an integrated scenario that addresses simultaneously
the variety of observed features.

%--------------------
\subsection{Outline}

%outline
In \se{shock_spheri} we provide an improved presentation of the original
analysis of spherical shock stability \citep[][hereafter BD03]{bd03}.
In \se{shock_cos} we compute the associated critical mass scale
in the cosmological context.
In \se{simu} we describe how the phenomena is demonstrated in cosmological
simulations, and learn about cold filaments in massive hot haloes at high
redshift.
In \se{feed} we address the cross-talk with the relevant feedback processes 
working alternatively below and above the critical scale.
Then, in \se{bimo}, we integrate the above processes into a scenario 
which attempts to explain the origin of the bi-modality and related features,
and report first results from simulations that implement the new ingredients.
In \se{discussion} we briefly discuss possible implications on other open
issues in galaxy formation,
and in \se{conc} we summarize our results, the proposed re-engineering of
galaxy formation simulations, and the open theoretical issues.

%2%%%%%%%%%%%%%%%%%%%%%%%%%%%%%%%
\section{Spherical Shock-Stability Analysis}
\label{sec:shock_spheri}

%D moved from old sec 2 simu
The standard paradigm of disc formation
\citep*{ro77,silk77,wr78,blum84,white91,mmw98},
which lies at the basis of all current models of galaxy formation,
assumes that while a DM halo relaxes to a virial equilibrium,
the gas that falls in within it is {\it shock heated\,}
near the halo virial radius $\Rv$ to the halo virial temperature.
It is then assumed to cool radiatively from the inside out.
As long as the cooling time is shorter than a certain global
free-fall time (or the Hubble time), typically inside a current
``cooling radius", the gas is assumed to accrete gradually onto a
central disc and then form stars in a quiescent way.
The maximum halo mass for efficient cooling was estimated to be on the
order of $\sim 10^{12-13}\msun$, and the common wisdom has been since
then that this explains the upper bound for disc galaxies.
However, early hints, based on one-dimensional simulations, indicated
that this scenario cannot reproduce the sharp drop in the luminosity
function above this scale \citep{thoul95}.
Even earlier studies, valid in the context of the pancake picture of galaxy
formation, indicated that virial shock heating may not be as general as assumed
\citep{binney77}. More advanced cosmological simulations have started to
reveal the presence of cold flows \citep{fardal01}.
With the new data from big surveys such as SDSS, 2MASS and 2dF,
and the detailed semi-analytic modeling (SAM) of galaxy formation,
it is becoming clear that the observed scale is somewhat smaller
and the drop is sharper than predicted by the original picture.
It seems that the current models based on the standard paradigm
have hard time trying to reproduce many of the observed bi-modality
features summarized in \se{intro}.
This motivated us to attempt a closer look at the shock-heating mechanism.

%-------------------------
\subsection{Spherical simulations}
\label{sec:simu_spheri}

\begin{figure}
\vskip 8.8cm
{\includegraphics{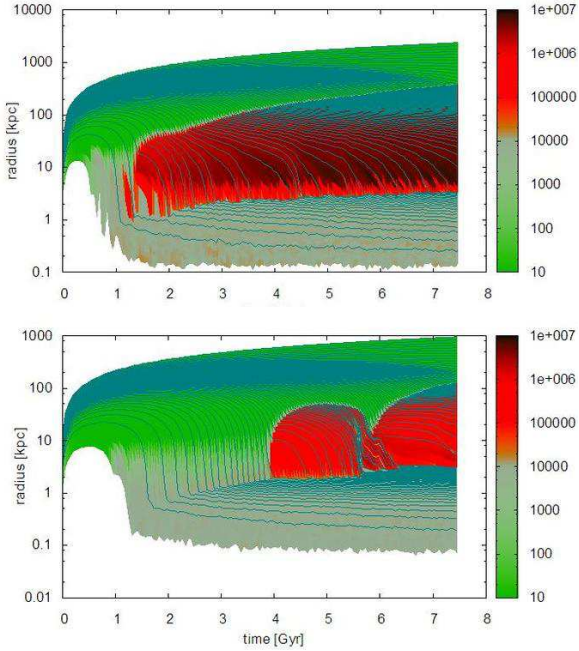}}
\vskip 0.2cm
\caption{Time evolution of the radii of Lagrangian
gas shells (lines) in a spherical simulation of a protogalaxy
consisting of primordial gas ($Z=0$) and dark matter.
Temperature is marked by color.
A shock shows up as a sharp break in the flow lines, namely a sudden slowdown
of the infall, associated with an abrupt increase in the temperature.
The lower discontinuity where the inflow is brought to a final halt
marks the ``disc" radius, formed due to an artificial centrifugal force.
(\textbf{a}) A massive system, where the virialized mass grows from
$10^{11}$ to $10^{13}\msun$.
(\textbf{b}) A less massive system, growing from $10^{10}$ to $10^{12}\msun$.
A virial shock exists only in systems more massive than a critical mass,
while in smaller haloes the gas flows cold and unperturbed into the
inner halo.
With more realistic metallicities the critical mass becomes
$\sim 10^{12}\msun$.
} 
\label{fig:bd03}
\end{figure}

\fig{bd03} shows the time evolution of the radii of Lagrangian gas shells
in a spherical gravitating system consisting of gas (in this case
with primordial composition) and dark matter, similar to the original
simulations by BD03 using an accurate one-dimensional hydrodynamical code.
Not shown are the dissipationless DM shells, which detach from
the cosmological expansion, collapse and then oscillate into virial equilibrium
such that they deepen the potential well attracting the dissipating gas shells.
The initial density perturbation was assumed to have a profile proportional
to the linear two-point correlation function of $\Lambda$CDM \citep[justified
in][]{dekel81},
and the final density profile roughly mimics the universal NFW profile
seen in cosmological simulations.
The gas is cooling radiatively based on the atomic cooling function computed
by \citet{sutherland93}.
The collapse of each gas shell is stopped at roughly $0.05 \Rv$
by an artificial centrifugal force which mimics the formation of a central
disc.

The upper panel focuses on massive haloes of $\sim 10^{12}\msun$.
As expected in the common picture, a strong shock exists near the virial
radius, namely at roughly half the maximum-expansion radius of the
corresponding shell. The virial shock gradually propagates outward,
encompassing more mass in time.  The hot post-shock gas is in a
quasi-static equilibrium, pressure supported against gravitational collapse.
The lower panel focuses on halo masses smaller by an order of magnitude,
and shows an interesting new phenomenon.
A stable shock forms and inflates from the disc toward the virial radius
only after a total mass of more than a few times $10^{11}\msun$ has collapsed.
In less massive systems, the cooling rate is faster than
the compression rate required for restoring the pressure in the
post-shock gas; had there been a shock, the post-shock gas would have
become unstable against radial gravitational contraction and unable
to support the shock.
In the specific case shown, with zero metallicity,
the critical mass is biased low; with more realistic metallicities
it becomes $\sim 10^{12}\msun$ due to the more efficient cooling via metal
lines (\se{shock_cos}).

% new
We demonstrate below (\se{simu}) that the behavior seen in the spherical 
simulations is robustly reproduced in cosmological simulations.
But first we wish to understand the spherical case, and use it
for simple predictions.

%------------------------------------------
\subsection{Post-shock stability criterion}
\label{sec:post}

The existence or absence of a shock, as seen in the simulations,
can be evaluated via a straightforward
stability analysis of the post-shock gas (first introduced in BD03).
We provide here a brief, improved presentation of this analysis,
followed by a more detailed estimate of the predicted critical mass
for shock stability as a function of redshift (\se{shock_cos}).

% SOME NEW WORDS
A stable extended shock can exist when the pressure in the post-shock gas 
is sufficient to balances the gravitational attraction toward the halo centre.
The standard equation of state for an ideal gas expresses the pressure as
$P=(\gamma-1)\rho\,e$
where $\rho$ and $e$ are the gas density and internal energy per unit mass,
and $\gamma =5/3$ for a mono-atomic gas.
In the text-book case of no cooling, the adiabatic index is defined as
$\gamma\equiv(\partial\ln P/\partial\ln\rho)_{\rm ad}$,
and the system is known to be gravitationally stable once $\gamma>4/3$.
When there is energy loss (e.g. by radiation) at a rate $q$ per unit mass,
we define a new quantity along the particle trajectories:
\be
\gamef \equiv {d \ln P \over d \ln \rho} 
= \gamma - {\rho \over \dot\rho}{q \over e}  \ .
\label{eq:gam}
\ee
The second equality follows from energy conservation, 
$\dot e = -P \dot{V} -q$ (where $V=1/\rho$),
plugged into the equation of state.
Note that $\gamef=\gamma$ when $q=0$.    
The difference between the two is a ratio of characteristic rates 
for the two competing processes: the cooling, which reduces the pressure 
in the post-shock gas, and the compression due to the pattern of the
post-shock infall, which tends to increase the pressure.
If the compression rate is efficient compared to the cooling-loss rate, it  
restores the pressure necessary for supporting a stable extended shock,
but otherwise the post-shock gas collapses inward under gravity, failing
to support the extended shock.

It is convenient to express the {\it compression rate\,} in the post-shock 
region as the inverse of a compression time, which we define by
\be
\tp \equiv \Gamma \frac{\rho}{\dot\rho} \ ,
\quad \Gamma \equiv \frac{3\gamma+2}{\gamma(3\gamma-4)}
= \frac{21}{5} \ ,
\label{eq:tp} 
\ee
with the factor $\Gamma$ to be justified below, and the last equality
referring to $\gamma=5/3$.
For a spherical shock at radius $\rs$, and a post-shock radial velocity $u_1$,
we assume that the radial flow pattern in the post-shock region is homologous, 
$u/r = u_1/r_s$. This is justified based on the 
spherical simulations described above, 
where the log-linear post-shock flow 
lines in \fig{bd03} are nearly parallel straight lines.
We then obtain using continuity
\be
\tp = \frac{\Gamma} { (-{\mathbf \nabla} \cdot {\mathbf u}) } 
    = \frac{\Gamma r_s}{(-3\, u_1)} \ .
\label{eq:tpu}
\ee

The competing {\it cooling rate\,} in the post-shock region is 
expressed as the inverse of the standard radiative cooling time defined by
\be
\tc \equiv \frac{e}{q} \ ,
\ee
where $e=e(T)$ and $q \prop \rho\, \Lambda(T,Z)$, functions of temperature $T$
and metallicity $Z$.
Then in \equ{gam}
\be
\gamef = \gamma - \Gamma^{-1} \frac{\tp}{\tc} \ .
\label{eq:gamef}
\ee

In order to test for stability, BD03 performed a perturbation analysis
were the radius of a shell is perturbed by 
$r \rightarrow r+\delta r$ 
and the sign of the force, $\ddot{\delta r} / \delta r$, is computed.
Writing $\delta r = u \delta t$, using the homology,
and assuming further that the gravity and pressure forces balance each other
near the transition state between stability and instability,
$\rho^{-1} \nabla P = GM/r^2$, 
one obtains a restoring force, i.e. stability, for 
\be
\gamef > \gamc \equiv \frac{2\gamma}{\gamma +2/3}
= \frac{10}{7} \ .
\label{eq:crit_gamma}
\ee
The $\gamc=10/7$ replaces the standard $\gamc=4/3$ of the adiabatic 
case.\footnote{
If the spherical symmetry assumed above is replaced by planar symmetry,
both for the shock and the gravitational field,
the stability criterion $\gamef > 10/7$ is replaced by
$\gamef > 10/11$ (Birnboim, Dekel \& Kravtsov, in preparation).
One can therefore assume in general that the actual critical value lies
somewhere between these two limits;
if $\gamef < 10/11$ there is no stable shock, if $\gamef > 10/7$ the conditions
allow a stable shock, and if $10/11<\gamef<10/7$ the shock stability
depends on the local geometry.
}

Using \equ{gamef} and the definitions of the time scales above,
the {\it shock stability criterion} of \equ{crit_gamma} becomes the simple
condition that {\it the cooling rate should be slower than the compression 
rate}:
\be
\tc > \tp \ .
\label{eq:criterion}
\ee
Once the cooling rate is slower, the pressure gain by compression can balance
the loss by radiative cooling, which allows the post-shock gas to be stable
against global gravitational collapse and thus support the shock.
The factor $\Gamma = 21/5$ has been introduced in the definition of 
$\tp$, \equ{tp}, in order to simplify this final expression.

% rates NEW
Note that the relevant quantity for stability is the {\it ratio\,} of rates
associated with the two competing processes, independent of how slow
each of them actually is in absolute terms. Each of the characteristic
times could in principle be longer than the Hubble time -- it is their ratio
which determines whether a stable shock is possible or the gas falls in
subject to gravity, cold and unperturbed.

%------------------
\subsection{Pre-shock quantities}
\label{sec:pre}
  
%------
\subsubsection{Compression rate}
\label{sec:pre_comp}

Using the standard jump conditions across a strong shock,
we can express the characteristic time scales (or $\gamef$) 
in terms of the pre-shock gas quantities. The jump condition
for the radial velocity is
\be
u_0-u_s = \frac{\gamma+1}{\gamma-1} (u_1-u_s) \ ,
\label{eq:jump_u}
\ee
where $u_s$ is the radial shock velocity 
and $u_0$ is the radial velocity of the pre-shock gas.  Then
\begin{eqnarray}
\tp &=& \frac{\Gamma (\gamma+1)}{3(\gamma-1)}
          \frac{\rs}{|u_0|}\,
          \left( 1 - \frac{2}{(\gamma-1)} \frac{\us}{|u_0|} \right)^{-1} \\
    &=&
          \frac{28}{5} 
          \frac{\rs}{|u_0|}\, 
          (1-3\ust)^{-1} \\
    &\simeq& 5.48\Gyr \, \frac{\rs}{|u_0|}\, (1-3\ust)^{-1} \ , % 5.4758
\label{eq:tpu0}
\end{eqnarray}
where $\ust \equiv \us/|u_0|$ (see \se{ust})
and the last expression assumes $\gamma=5/3$,
$\rs$ in $100\kpc$, and $u_0$ in $100\kms$.
If $u_s=0$, say,
then $\tp$ is about 6 times larger than $\rs/|u_0|$, a typical
free-fall time from $\rs$ into the halo centre. At the virial radius,
$\tp$ is comparable to the Hubble time at the corresponding epoch,
but at inner radii it becomes significantly shorter.

%------
\subsubsection{Cooling rate}
\label{sec:pre_cool}

The cooling time \citep[e.g. based on][]{sutherland93} is
\be 
\tc \equiv \frac{e}{q}= \frac{(1+2\epsilon)}{(1+\epsilon)} \frac{3}{2}kT
\left[\frac{\chi^2}{m}\, \rho\, \Lambda(T,Z) \right] ^{-1} \ ,
\ee
where $\Lambda(T,Z)$ is the cooling function,
$k$ is the Boltzmann constant, $\eps \equiv n_{He}/n_H$,
the mass per particle is $m \equiv \mu m_p$ with $\mu=(1+4\eps)/(2+3\eps)$,
and the number of electrons per particle is $\chi=(1+2\eps)/(2+3\eps)$.
For 25\% He in mass, one has $\eps=1/12$, yielding $\mu=0.59$.
If we express the post-shock temperature as $T_6 \equiv T/10^6K$,
the post-shock baryon density as 
$\rho_{-28} \equiv \rho/10^{-28} {\rm g}\, {\rm cm}^{-3}$,
and the cooling function as $\Lambda_{-22}(T,Z) \equiv \Lambda(T,Z)/10^{-22}
{\rm erg}\, {\rm cm}^3\, {\rm s}^{-1}$,
we have
\be
\tc \simeq 2.61\Gyr\, \rho_{-28}^{-1}\, T_6\, \Lambda_{-22}^{-1}(T,Z)\ .%2.6059
\label{eq:tcu0}
\ee

The post-shock gas density is related to the pre-shock density by
the jump condition
\be
\rho_1 =\frac{\gamma+1}{\gamma-1}\rho_0 = 4\rho_0  \ ,
\label{eq:rho_jump}
\ee
and the post-shock temperature entering the cooling time
is related to the pre-shock radial velocity $u_0$ via
\be
\frac{k T_1}{m} = \frac {2(\gamma -1)}{(\gamma+1)^2}\,  (u_0-u_s)^2 
= \frac{3}{16}\, u_0^2\, (1+\ust)^2\ .
\label{eq:T1u0}
\ee

We note in passing that for a virial shock, where $u_0=-\Vv$ (BD03),
the post-shock temperature is actually
\be
T_1 \gsim \frac{3}{8}\, \Tv \ ,
\ee 
comparable to but somewhat smaller than the virial temperature as defined 
in \equ{Tv}.

%-----
\subsubsection{Stability criterion}
\label{sec:pre_crit}

Using eqs.~\ref{eq:tpu0} and \ref{eq:tcu0}, 
the critical stability condition becomes
\be 
\frac{\tc}{\tp} 
\simeq 0.48\, \frac{\rho_{-28}^{-1} T_6\, \Lambda_{-22}^{-1}(T,Z)} % 0.476
                   {\rs\, |u_0|^{-1}\, (1-3\ust)^{-1}}
\simeq 1 \ ,
\label{eq:crit}
\ee
with $\rs$ in $100\kpc$ and $u_0$ in $100\kms$.
Recall that \equ{rho_jump} relates $\rho$ to $\rho_0$, and 
\equ{T1u0} relates $T$ to $u_0$.
Thus, for given shock radius $\rs$, shock velocity relative to infall 
$\us/|u_0|$, and pre-shock gas density $\rho_0$, 
once the metallicity $Z$ is given and the cooling function 
$\Lambda(T,Z)$ is known, one can solve \equ{crit} for
the critical values of $T$ and the corresponding $u_0$.
When put in a cosmological context (\se{shock_cos}), 
this solution is associated with a unique critical halo mass. 

% stability criterion in spherical simulations
The stability criterion derived above, \equ{crit_gamma} or \equ{criterion},
is found to work very well when compared to the results of the spherical 
simulations shown in \fig{bd03}.  When $\gamef$ (or $\tc/\tp$) 
is computed using pre-shock quantities at a position 
just outside the ``disc", we find that
as long as the halo is less massive than a critical scale, 
before the shock forms, the value of
$\gamef$ is indeed well below $\gamc$ and is gradually 
rising, reaching $\gamc$ almost exactly when the shock starts propagating
outward. The value of the $\gamef$ computed using the quantities just  
outside the shock then oscillates about $\gamc$ with a decreasing amplitude,
following the oscillations in the shock radius seen in \fig{bd03}.
As the shock eventually settles at the virial radius, 
$\gamef$ approaches $5/3$, larger than $\gamc=10/7$,
where the cooling is negligible.
% Stability criterion in cosmological simulations
The same stability criterion is found to be valid to a good approximation
also in cosmological simulations (\se{simu}).

%-------------------
\subsubsection{Shock velocity}
\label{sec:ust}

What value of $\ust$ is relevant for evaluating stability?
In the inner halo, we use $\ust=0$.
This is because, as the halo is growing in mass,
the shock first forms in the inner halo and then propagates outward
(\fig{bd03}b). The onset of shock stability is therefore marked by
its ability to develop a velocity outward.

During the stable phase when the shock is expanding with the virial radius,
the spherical simulations indicate roughly $u_1 \simeq -\us$ (\fig{bd03}a),
namely $\ust \simeq 1/7$ (\equnp{jump_u}). 
This indicates that a small shock velocity of such a magnitude is appropriate
for evaluating stability at the virial radius.

Note that stability is harder to achieve when the shock is expanding
relatively fast. In particular, in the extreme case $\ust = 1/3$, 
the post-shock velocity vanishes, $u_1=0$ (\equnp{jump_u}). 
The compression rate becomes infinitely slow (\equnp{tpu}),
implying that the shock cannot be stabilized.

%%%%%%%%%%%%%%%%%%%%%%%%%%%%%%
\section{Shock-Heating Scale in Cosmology}
\label{sec:shock_cos}

%-------------------------
\subsection{Haloes in cosmology}

We wish to translate the critical stability condition, \equ{criterion} or
\equ{crit}, into a critical post-shock temperature, and the corresponding
critical halo virial velocity and mass as a function of redshift.
\Equ{crit} has a unique solution when combined with
the two virial relations between halo mass, velocity and radius
(\equnp{virial}), 
and the relation between post-shock temperature and pre-shock infall velocity
(\equnp{T1u0}).

As summarized in Appendix A, the time dependence of the virial relations 
can be expressed in terms of the convenient parameter
\be
A\equiv (\Delta_{200} {\omm}_{0.3} h_{0.7}^2)^{-1/3}\, a\ ,
\ee
where $a\equiv 1/(1+z)$ is the cosmological expansion factor
and the other parameters are of order unity.
The parameters ${\omm}_{0.3}$ and $h_{0.7}$ correspond to today's values
of the cosmological mass density parameter and the Hubble expansion parameter
respectively, and for the standard $\Lambda$CDM cosmology adopted here they 
are both equal to unity.
The parameter $\Delta_{200}$ is the virial density factor given approximately 
in \equ{Delta}: at redshifts $z>1$ it is $\Delta_{200} \simeq 1$,
but at lower redshifts it becomes somewhat larger, reaching 
$\Delta_{200} \simeq 1.7$ at $z=0$.

%-------------------------------
\subsection{Compression rate}

% t_p
For a shock at the virial radius,
$r_s=\Rv$, we approximate $u_0=-\Vv$, as predicted 
by the spherical collapse model in an Einstein-de Sitter cosmology
(BD03, Appendix B).

When the shock is at an arbitrary inner radius $r$, 
where the infall velocity is $|u|$,
we multiply $\Rv$ and $\Vv$ by appropriate
factors $\fr\equiv r/\Rv$ and $\fu\equiv |u|/\Vv$ (discussed in \se{disc}).
Then \equ{tpu0} becomes
\be
\tp\simeq 14.3\Gyr\, A^{3/2}\, \fr \fu^{-1} ( 1-3\ust)^{-1} \ . %14.282
%F_{\rm p} \equiv \fr \fu^{-1} ( 1-3\ust)^{-1} .
\label{eq:tpc}
\ee

%-------------------------------------
\subsection{Cooling rate: gas density}

%t_c
In order to express the cooling time of \equ{tcu0}
in terms of cosmological quantities, we first evaluate the 
pre-shock baryon density, which we write as 
\be
\rho_{\rm b}= 4\, \fb\, (\rho/\bar\rho)_{\rm vir}\, 
\Delta\, \rhou\, f_\rho \ .
\label{eq:rhob}
\ee
Here $\rhou$ is the universal mean mass density (\equnp{rhou}),
and $\Delta$ is the top-hat mean overdensity inside the virial radius 
(\equnp{Delta}).
The factor $(\rho/\bar\rho)_{\rm vir}$ translates $\bar\rho$,
the mean total density interior to $\Rv$, to $\rho$, the local total 
density at $\Rv$.
The effective baryonic fraction $\fb$ turns it into a pre-shock
baryonic density.
The factor $4$ stands for the ratio between the post-shock gas density
and the pre-shock gas density.\footnote{In 
the spherical simulations, the relevant factor relating the baryon
density to the DM density in \equ{rhob}
is actually closer to $\sim 3$
because of a ``bump" in the DM density just inside the virial radius.}
The factor $f_\rho\equiv \rho(r)/\rho(\Rv)$ reflects the ratio of the actual 
gas density at some radius $r$ within the halo to its value at the virial 
radius (see \se{disc}).

The ratio $\rho/\bar\rho$ at the virial radius is derived for the universal
NFW halo density profile revealed by cosmological simulations 
\citep*{nfw97}.  For a virial concentration parameter $c$, this ratio is
\be
\left( \frac{\rho}{\bar\rho} \right) _{\rm vir}
=\frac{c^2}{3(1+c)^2} \left[ \ln(1+c)-\frac{c}{(1+c)} \right]^{-1} .
\ee
A typical concentration of $c=12$ is associated with 
$\rho/\bar\rho \simeq 0.17$; we therefore express the approximate results 
below using the factor $f_{\bar\rho,0.17} \equiv (\rho/\bar\rho)/0.17$.
In our more accurate evaluation of the critical scale (\se{accurate}),
we model the dependence of the mean concentration on mass and time using the
fit of \citet{bullock01_c} for the $\Lambda$CDM cosmology:
\be
c(M,a) = 18\, M_{11}^{-0.13}\, a \ .
% \simeq 9.5 \left(\frac{M}{M_*(a)} \right) ^{-0.13}.
\label{eq:conc}
\ee

The effective baryon fraction $\fb$ may in principle be as large as the
universal fraction $\simeq 0.13$, but it is likely to be smaller 
because of gas loss due to outflows.
For the approximate expressions we define $f_{{\rm b},0.05} \equiv \fb/0.05$.

The gas density at $r$, \equ{rhob}, thus becomes 
\be
\rho_{-28} =0.190\, A^{-3}\, f_{{\rm b},.05}\, f_\rho\, 
f_{\bar\rho,0.17} \ .   % 0.19041
\label{eq:rho28}
\ee
Inserting this baryon density into \equ{tcu0}, 
the cooling time becomes
\be
\tc=13.7\Gyr\, A^3 f_{b,.05}^{-1} f_\rho^{-1} f_{\bar\rho,0.17}^{-1}\, 
     T_6\, \Lambda_{-22}^{-1}(T,Z) .  % 13.685
\label{eq:tcrhob}
\ee

% CMB and reionization
The cooling function that we use below \citep[based on][]{sutherland93}
neglects two physical processes:
Compton scattering off the cosmic microwave background
and the possible effect of external radiation on the cooling rate
through the reionization of Hydrogen.
Based on the more complete cooling function as implemented by 
\citet{krav04}, one learns that
these processes become important only for densities below $\sim 10^{-28}$ and 
$\sim 10^{-26}\, {\rm g}\, {\rm cm}^{-3}$ at $z\sim 0$ and $4$ respectively.
Using \equ{rho28}, we conclude that while these processes may have a certain 
effect on the cooling rate near the virial radius, they should be negligible 
once the analysis is applied inside the inner half of the halo, where 
the critical scale for shock heating is determined in practice.
We address these effects in more detail elsewhere 
(Birnboim, Dekel \& Loeb, in preparation).
%At Rv, for z=0 and 4, rho~3.4x10^-29 and 2.5x10^{-27} respectively.

%----
\subsection{Metallicity}

The metallicity near the virial radius and in the inner halo, which also
enters the cooling rate, is one of our most uncertain inputs.
For the mean metallicity $Z$ (in solar units) as a function of redshift $z$
we use the two-parameter functional form
\be
\log (Z/Z_0)=  -s\,z \ ,
\label{eq:metal}
\ee
where $Z_0$ is today's metallicity and the slope $s$ governs the rate of 
growth.

An upper limit may be imposed by the hot, X-ray emitting 
Intra-Cluster Medium (ICM) at low redshifts, which indicate $Z_0 \sim 0.2-0.3$.
The ICM metallicity evolution in Semi-Analytic Models, assuming a range of
different feedback recipes, yields consistently an average enrichment rate of
roughly $s \simeq 0.17$  
\citep*[R. Somerville, private communication;][]{delucia04_enrich}.
We adopt this enrichment rate $s$ in our modeling below.

A realistic estimate of the metallicity near the virial radius 
(or perhaps a lower limit for the inner halo) 
may be provided by C$_{IV}$ absorbers in the Inter-Galactic Medium (IGM) at 
$z \sim 2-4$ \citep{schaye03}. 
At densities appropriate to typical NFW haloes at $z=3$ (with $c=3$), 
namely $\rho_{\rm vir} \simeq 53\rhou$, 
they measure an average of $[C/H]=-2.47$. 
Silicon measurements, on the other hand, seem to indicate a metallicity 
that is about five times larger (A. Aguirre, private communication). 
If one takes the geometrical mean between the metallicities 
indicated by C$_{IV}$ and by Si one has $Z(z=3) \simeq 0.0075$.
This translates to $Z_0=0.025$ if $s=0.17$. 

We note that another popular indicator, Mg$_{II}$, indicates consistently 
$Z<0.01$ within $50-100\kpc$ of galaxy centres at $z \sim 1$
\citep[private communication with J. Charlton; e.g.][]{ding03}.
%masiero04 to be submitted

The Damped Lyman-Alpha Systems (DLAS) are believed to sample cold gas
deeper inside the haloes, and can thus provide another interesting limit.
Observations in the range $z=1-4$ \citep{prochaska03}
can be fitted on average by \equ{metal} with $Z_0 \simeq 0.2$ and 
a somewhat steeper slope $s \simeq 0.26$. However,
a fit with $s=0.17$ (and then $Z_0=0.1$) is not ruled out. 

Based on the above estimates, we adopt as our fiducial metallicities
$Z_0=0.03$ at $\Rv$ and $Z_0=0.1$ at the ``disc" radius $\sim 0.1\Rv$, 
both with an enrichment rate $s=0.17$.

%-----------------------------
\subsection{Inside the halo}
\label{sec:disc}

For a shock in the inner halo we wish to estimate the factors
$f_r$, $f_u$ and $f_\rho$.

Empirically from the spherical simulation of BD03,
for a shell encompassing a mass just shy of the critical mass
(as well as from the toy model of BD03 of gas contracting in a static 
isothermal sphere), we estimate for $f_r\equiv r/\Rv$
\be
f_\rho \equiv \frac{\rho_0(r)}{\rho_0(\Rv)}
\simeq
\cases{&$\!\!\!\!\!f_r^{-1.6},\quad r\lsim \Rv$\cr
       &$\!\!\!\!\!f_r^{-2.1},\quad r\sim 0.1\Rv$ } .
\label{eq:tp1}
\ee
We adopt below $f_\rho = f_r^{-2}$ at $f_r=0.1$.

Energy conservation assuming pure radial motion inside a static singular
isothermal sphere yields
\be
f_u\equiv \frac{u_0(r)}{u_0(\Rv)}
=[1 +2 \fb (f_r^{-1} -1) +2(1-\fb)\ln f_r^{-1}]^{1/2} .
\ee
For $f_r=0.1$ and $\fb=0.05$ this gives the estimate $f_u \simeq 2.5$.

Based on the definition of $f_u$,
the temperature behind a virial shock is related to the temperature
obtained from the stability condition at radius $r$ by
\be
T(\Rv) = f_u^{-2}\, T(r) \ .
\label{eq:T1}
\ee

%----------------------------
\subsection{Crude explicit estimates}
\label{sec:approx}

The critical temperature for stability, as obtained by comparing $\tc$
and $\tp$ in the cosmological context, \equ{tcrhob} and \equ{tpc}, is
\be
T_6\, \lambda_{-22}^{-1} (T,Z) = 1.04\, A^{-3/2}\, F \ , % 1.0436
\label{eq:t6}
\ee
where
\be
F \equiv f_r f_u^{-1} f_\rho \ f_{{\rm b},.05} f_{\bar\rho,0.17}
(1-3\ust)^{-1} \ .
\ee

%\subsubsection{An approximate cooling function}

The cooling function as computed by 
\citet{sutherland93} 
can be crudely approximated in the range $0.1 < T_6 < 10$ by
\be
\Lambda_{-22} \simeq 0.12\, Z_{0.03}^{0.7} \, T_6 ^{-1} + 0.02\, T_6 ^{1/2} \ ,
\label{eq:lamfit}
\ee
where $Z_{0.03} \equiv Z/0.03$, with $Z$ in solar units.
The above expression is valid for $-2.5 \leq \log Z \leq 0$, and at 
lower metallicities the value of $\Lambda$ is practically the same as for 
$\log Z = -2.5$.
This fit is good near $T_6 \sim 1$ for all values of $Z$.
The first term refers to atomic cooling, while the second term
is due to Bremsstrahlung.  For an approximation relevant in 
haloes near $\msh$ we ignore the 
Bremsstrahlung term, which becomes noticeable only at higher temperatures.
One can then obtain in \equ{t6} an analytic estimate for
the critical temperature:
\be
T_6 \simeq 0.36\, A^{-3/4}\, (Z_{0.03}^{0.7} F)^{1/2}  \ . %0.3539
\label{eq:tapp}
\ee

Using \equ{T1u0} and \equ{T1}, with $|u_0|=\Vv$ just outside the virial radius,
we then obtain for the critical velocity and mass
\be
V_{100} \simeq 1.62 A^{-3/8}\, 
(Z_{0.03}^{0.7} F)^{1/4}\, \fu^{-1}\, (1+\ust)^{-1} \ , % 1.6217
\label{eq:vapp}
\ee
\be 
M_{11} \simeq 25.9 A^{3/8}\,
(Z_{0.03}^{0.7} F)^{3/4}\, \fu^{-3}\, (1+\ust)^{-3} \ .  % 25.86
\label{eq:mapp}
\ee

% ust
A comment regarding the $\ust$ dependence of our results.
The critical temperature depends on the shock velocity $\ust$ via $F$, 
$T \prop (1-3\ust)^{-1/2}$, reflecting the $\ust$ dependence of $\tp$.
The critical temperature is thus monotonically increasing with $\ust$.
An additional $\ust$ dependence enters when the temperature is
translated to a critical virial velocity using the jump condition, 
$V \prop (1+\ust)^{-1} T^{1/2}$, and then to a critical mass, 
$M \prop (1+\ust)^{-3} T^{3/2}$.
For a slowly moving shock, $\ust \ll 1/3$, 
the combined $\ust$ dependence of the critical mass is 
$M \prop [1+(9/4)\ust](1-3\ust) \simeq 1-(3/4)\ust$ --- 
a decreasing function of $\ust$. 
This means that at a given radius in a halo of a given mass, 
when everything else is equal, 
a slowly expanding shock is actually more stable than a shock at rest.
For example, if the shock is expanding with $\ust=1/7$ rather
than $\ust=0$, the critical mass is smaller by about 24\%.
However, recall that stability is harder to achieve when the shock is expanding
relatively fast, and the compression completely vanishes if $\ust \geq 1/3$ 
(\se{ust}).

%\subsubsection{The virial radius versus the inner halo}

For actual crude estimates of the critical scales at $z=0$, 
we assume $f_{{\rm b},0.05} \simeq f_{\bar\rho,0.17} \simeq 1$.
For a shock at the {\it virial radius},
$f_r=f_u=f_\rho=1$, we assume $Z_0 \simeq 0.03$ and $\ust \simeq 1/7$,
and obtain
\be
T_6 \simeq 0.5, \quad 
V_{100}\simeq 1.6, \quad  
M_{11}\simeq 26. 
\ee

At an {\it inner radius\,} closer to the disc vicinity, say $f_r =0.1$, 
we estimate $f_u \simeq 2.5$ and $f_\rho \simeq 100$ (\se{disc}).
Assuming $Z_0 \simeq 0.1$ and $\ust \simeq 0$,  we obtain 
\be
T_6 \simeq 1.1, \quad   %1.082 
V_{100}\simeq 1.1, \quad  %1.1363
M_{11}\simeq 8.8.   %8.806 
\ee
We see that for a shock at $r\sim 0.1\Rv$, 
the expected critical mass is smaller than at $\Rv$, 
somewhat below $\sim 10^{12}\msun$.

The above estimates are useful for exploring the qualitative dependences of
the critical values on redshift, metallicity and gas fraction.
For example, in \equ{mapp},
the explicit redshift dependence and the decrease of
metallicity with redshift tend to lower the critical mass toward higher $z$. 
On the other hand, the decrease of halo concentration with $z$ (i.e. increase 
of $f_{\bar\rho,0.17}$),
and the possible increase of the effective gas fraction with $z$ 
(\se{bimo}), tend to push the critical mass up at higher $z$.

%----------------------------
\subsection{More accurate estimates}
\label{sec:accurate}

We now obtain a better estimate of the critical temperature 
(and then critical mass and virial velocity)
by solving \equ{t6} using the exact cooling function of SD93 
and adopting specific models for the time evolution of metallicity 
and halo structure. The results are presented graphically.
  
The baryon density is computed assuming an NFW profile whose concentration
evolves in time as in \equ{conc}.
The effective fraction of cold gas is assumed to be $\fb=0.05$, 
motivated by best fits of semi-analytic models to the 
Milky Way 
\citep*{klypin02} 
and by fitting the $\Lambda$CDM halo mass function
to the observed luminosity function near $L_*$ 
\citep{bell03_sam}.
The metallicity evolution is parametrized as in \equ{metal} with $s=0.17$
for today's metallicities in the range $Z_0=0.03-0.3$.
Upper and lower estimates for the critical scales
are evaluated at the virial radius and at $r=0.1\Rv$ respectively,
using the crude estimates of \se{disc}.
In the following figures the shock is assumed to be at rest, $\us=0$.

%4
\begin{figure}
\vskip 7.2cm
{\includegraphics{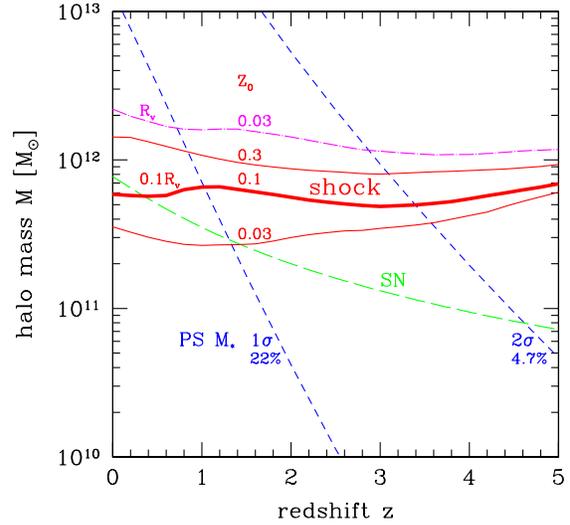}}
\caption{
Critical shock-heating halo mass as a function of redshift. 
The three solid (red) curves refer to a shock at the inner halo, $r=0.1\,\Rv$,
with different metallicities as indicated; the middle curve ($Z_0=0.1$) is our
best estimate.
The dash-dotted (magenta) curve refers to a shock at the virial radius 
with $Z_0=0.03$.
The other parameters used are: $\fb=0.05$, $\us=0.$, $s=0.17$ (see text).
Shown for comparison (short dash, blue) are the Press-Schechter estimates
of the forming halo masses, corresponding to 1-$\sigma$ ($\mps$) 
and 2-$\sigma$, where the fractions of total mass in more massive haloes
are 22\% and 4.7\% respectively.
Also shown is the critical mass for supernova feedback 
discussed in \se{feed} (long dash, green).
}
\label{fig:coolzm}
\end{figure}

%5
\begin{figure}
\vskip 7.2cm
{\includegraphics{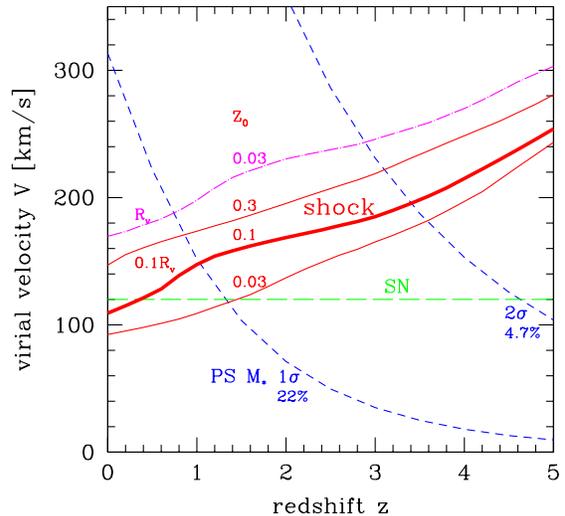}}
\caption{
Same as \fig{coolzm} but for the corresponding halo virial velocity.
}
\label{fig:coolzv}
\end{figure}

%6
\begin{figure}
\vskip 7.2cm
{\includegraphics{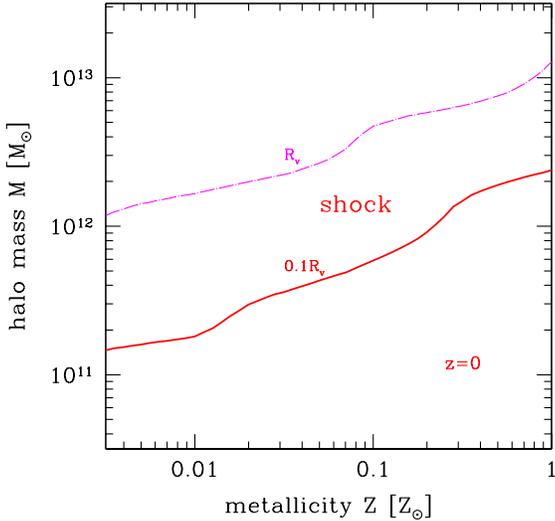}}
\caption{
Critical shock-heating halo mass as a function of metallicity at redshift 
$z=0$.  The solid (red) curve refers to a shock at the inner halo,
$r=0.1\,\Rv$. The dash-dotted (magenta) curve refers to a shock at the virial
radius.
}
\label{fig:coolzz}
\end{figure}

\Fig{coolzm} shows the critical mass 
as a function of redshift.  At a typical inner-halo radius, $r=0.1\,\Rv$,
we plot the curves for three different current metallicities:
$Z_0=0.03,\,0.1,\,0.3$.
The critical halo mass, for $Z_0=0.1$,
is $\simeq 6\times 10^{11}\msun$ quite independent of redshift. 
The uncertain metallicity introduces a scatter by
a factor of 2 up and down (for $z<2.5$).

An upper limit of $\sim 2\times 10^{12}\msun$ is obtained for a shock at $\Rv$
when a correspondingly low metallicity is assumed, $Z_0=0.03$.
When the assumption of $\ust \simeq 0$ is replaced by 
$\ust \simeq 1/7$, allowing the shock to expand with the virial radius 
as seen in \fig{bd03}a, the critical mass at $\Rv$ with $Z_0=0.03$ 
becomes comparable to that at $0.1\,\Rv$ with $Z_0=0.3$.

\Fig{coolzv} shows the corresponding virial velocity.
At $z=0$ the critical virial velocity for a shock in the inner halo
is $\sim 120\kms$, with a $\pm 30\kms$ scatter due to metallicity. 
The critical virial velocity increases monotonically with redshift, to 
$\sim 200\kms$ near $z\sim 3$ (a crude fit to the redshift dependence is
$\Vv=120+28z$).

% coolzz
The dependence on metallicity at $z=0$ is highlighted in \fig{coolzz}.
The metallicity enters strongly through the cooling function $\Lambda(T,Z)$. 
The critical mass grows roughly like $Z^{1/2}$, as predicted in \equ{mapp},
so it spans about an order of magnitude over the whole metallicity range.

% app
The analytic estimates of \equ{vapp} and \equ{mapp},
based on the approximate cooling function,
turn out to provide good estimates in most cases, and can therefore be used
for extending the results analytically to any desired choice of the relevant
parameters.

% disc vs vir
We learn that the critical halo mass for shock stability at the disc vicinity,
$\msh(r_{\rm disc})$, 
is somewhat smaller than for a shock at the virial radius, $\msh(\Rv)$.
This result is robust: it is true even if the metallicity at the virial 
radius is smaller by an order of magnitude than the metallicity at the disc,
and even when $\ust$ at $\Rv$ is as large as $1/7$.
This means that as the halo is growing in mass, the conditions for a stable 
shock develop first in the inner halo and somewhat later in the outer halo.
Thus, in haloes of mass $M<\msh(r_{\rm disc})$, 
we expect cold flows with no shock heating throughout the halo.
In the other extreme of haloes of mass $M> \msh(\Rv)$, we expect shock heating
of most of the gas by a shock near the virial radius.
In haloes of mass in the narrow intermediate range 
$\msh(r_{\rm disc}) < M < \msh(\Rv)$,
we expect shock heating somewhere inside the halo, preventing most of the
gas from falling in and giving rise to a hot medium.
This predicted mass range, of a factor of 2-3 in mass,
is consistent with the range of transition from all cold to mostly hot 
seen in cosmological simulations (\se{simu}).
%We discuss below (\se{streams}) the appearance of cold streams 
%along the denser filaments in haloes in this regime.

% M*
Also shown in \fig{coolzm} are the typical masses of haloes forming
at different redshifts, the 1-$\sigma$ (termed $\mps$) and 2-$\sigma$
halo masses according to the Press-Schechter formalism, \equ{mstar}.
According to the improved Sheth-Tormen version, the corresponding fractions
of the total mass encompassed in haloes exceeding the mass $M$
are 22\% and 4.7\% respectively.
One can see in \fig{coolzm} that $\msh$ coincides with $\mps$ at $z \sim 1$, 
and with the 2-$\sigma$ mass at $z\sim 3.4$. 
By $z\sim 2$, say, most of the forming haloes are significantly less massive
than $\msh$.
When embedded in a large-scale high-$\sigma$ density peak, the
distribution of forming haloes at a given $z$ may shift toward more 
massive haloes. In fact, the most massive halo in a volume of comoving size
$\sim 100 {\rm Mpc}$ is likely to be more massive than $10^{12}\msun$
at all relevant redshifts ($z < 6$, say).
Nevertheless, the qualitative result concerning the majority of the haloes
remains valid. 
We conclude that {\it in the vast majority of forming galaxies
the gas has never been shock-heated to the virial temperature\,} 
-- it rather flows {\it cold\,} all the way to the disc vicinity.

% comparable to obs
We note that the values obtained for $\msh$ at low redshifts 
are compatible with the observed bi-modality/transition scale summarized 
in \se{intro}.
The estimates in the inner halo, using the lower and upper limits for
$Z_0$, indeed border the observed characteristic
halo mass of $\sim 6\times 10^{11}\msun$.
The upper-limit estimate at $\Rv$ corresponds to a halo mass similar to 
that of the Milky Way.

%%%%%%%%%%%%%%%%%%%%%%%%%%%%%%%%%%%%%%%%%%%%%%%%%%%%
\section{Cold Streams in Hot Haloes}
\label{sec:simu}

%----------------------------------
\subsection{Cosmological simulations}

\begin{figure}
\vskip 10.4cm
{\includegraphics{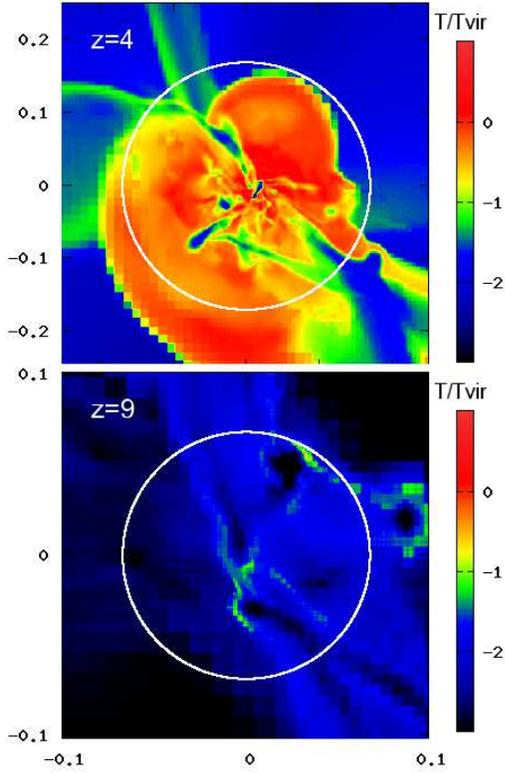}}
\vskip 0.2cm
\caption{Snapshots from a cosmological hydrodynamical simulation
(Birnboim, Dekel, Kravtsov \& Zinger, in preparation, hereafter BDKZ05) %
showing the gas temperature in a slice 
of a protogalaxy at two different epochs, when it has two different masses.
The temperature is relative to the virial temperature of the halo
at that time.  The side of each slice is scaled to be $3\Rv$ 
(numbers in comoving $\!\hmpc$).
{\bf Top:} At $z \simeq 4$, when the halo is already relatively massive,
$M \simeq 3\times 10^{11}\msun$.
Much of the gas is heated by a strong shock near the virial radius
(white circle). 
Cold streams penetrate through the hot medium deep into the halo.
{\bf Bottom:} At $z \simeq 9$, when the halo is still small,
$M \simeq 2\times 10^{10}\msun$.
The gas flows in cold ($T \ll \Tv$), showing no evidence for
shock heating inside the virial radius (circle).
}
\label{fig:krav}
\end{figure}

% from old simu section, before the analysis
Cosmological hydro simulations indicate that the phenomenon of cold 
flows is a general phenomenon not restricted to spherical symmetry. 
\fig{krav} displays snapshots of an Eulerian simulation
from Birnboim, Dekel, Kravtsov \& Zinger (in preparation, 
BDKZ05).\footnote{A description of the simulation technique can be found in
\citet{krav03},
% or \citet{krav04}, 
where it was used for other purposes.}
Shown are maps of gas temperature in two epochs in the evolution of
a protogalaxy:
one at $z\simeq 4$, when the halo is already relative massive,
and the other at $z \simeq 9$, when the halo is still rather small.
While the more massive halo, near the critical scale,
shows a hot gas component near the virial
temperature behind a virial shock, the smaller halo shows
only cold flows inside the virial radius.

Similar results have been obtained earlier from SPH simulations
by \citet{fardal01}, who emphasized the feeding of galaxies by cold
flows preferentially at early epochs. Based on our spherical analysis,
we understand that this redshift dependence mostly
reflects the smaller masses of haloes at higher redshifts.
\citet{keres04} have analyzed similar SPH simulations and presented
the case for the two modes of infall, cold and hot, in more detail.\footnote{
The hot phase becomes an ``infall" mode in this simulation after the gas 
cools, but in reality it may be kept hot and be prevented from falling in by
feedback effects, \se{feed}.}  
For example, in their figures 1 and 2 they demonstrate the two-mode
phenomenon very convincingly by showing the distribution of particles and their
trajectories in temperature-density diagrams.
Their most informative Fig.~6 
shows the fractions of cold and hot infall as a function of halo mass
at different redshifts. For all haloes of masses 
below a critical mass the infall is predominantly cold. Near the critical
mass there is a relatively sharp transition into a hot mode, which becomes
dominant above the critical mass. 
The transition from 100\% cold to more then 50\% hot occurs across
a range of only a factor $\sim 2$ in halo mass.
In this simulation, where zero metallicity is
assumed, the transition mass is $M \sim 3\times 10^{11}\msun$ at all
redshifts in the range $0-3$. 
A similar transition as a function of halo mass, and the constancy
of the critical mass as a function of redshift, are both reproduced
in the Eulerian cosmological simulations studies in BDKZ05.

The spherical simulations described above (\se{simu_spheri}), 
and the corresponding analytic analysis (\se{shock_spheri}, \se{shock_cos}),
yield very similar results. In fact, our analytic predictions
for the case of zero metallicity match the critical mass measured by
\citet{keres04} remarkably well. 

%stability criterion and cosmological simu (from the analysis section)
The stability criterion derived in the spherical case turns out to be valid
locally in the cosmological simulations where the non-spherical features
are pronounced.
BDKZ05 use this criterion to identify the cold streams and hot media
in the simulations without explicit information concerning the presence 
or absence of actual shocks.
When testing the criterion in these simulations, 
in which the hot and cold phases may be present in the same halo,
the local gas properties at each position is first transformed to
post-shock quantities, as if there was a shock there, and the 
stability is evaluated based on the derived value of $\gamef$ there.
The resultant maps of $\gamef$ resemble quite well the temperature maps
of the actual simulation.
This demonstrates that the wisdom gained by the spherical analysis
is applicable in the general case.

%------------------------------
\subsection{Filaments in the simulations}

\begin{figure}
\vskip 6.1cm
{\includegraphics{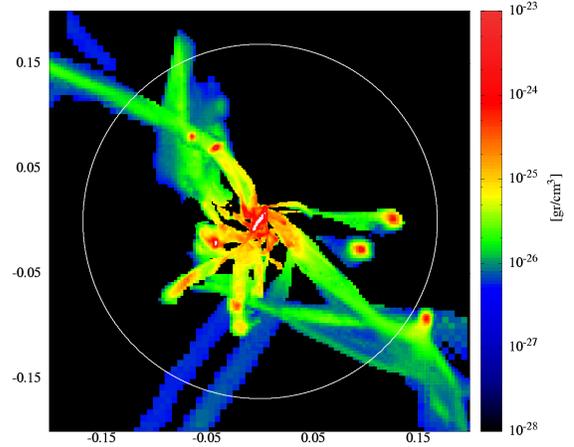}}
\vskip 0.2cm
\caption{A snapshot from a cosmological hydrodynamical simulation  
(Zinger, Birnboim, Dekel \& Kravtsov, in preparation, ZBDK05) 
showing the gas density of the cold flows 
($T<10^{4.5}$K, $\rho>10^{-26.3} g\,cm^{-3}$) 
within the virial radius of the same massive galaxy shown in \fig{krav} 
at $z=4$.
The cold phase is filamentary. In the outer radii, the gas filaments 
tend to ride on the large-scale DM filaments feeding the halo.
About half the mass of the cold phase is in dense clumps.
}
\label{fig:zinger}
\end{figure}

The cosmological simulation of \fig{krav} also reveal that massive haloes at 
high redshift can have cold streams embedded in the hot medium.  
While the hot medium is rather spherical, the cold flows are {\it filamentary}
and sometimes clumpy.
\fig{zinger} highlights the filamentary nature of the cold gas in the
same big halo. 
This phenomenon is consistent with the findings of \citet{keres04}, that
the cold mode may in some cases co-exist with the hot mode 
even above $\msh$, especially at $z>2$ (their Fig.~6)
and preferentially in relatively isolated galaxies.
We wish to understand the origin of this phenomenon, and learn about 
its dependence on cosmological time.

The simulation results of \citet{keres04} provide several additional 
relevant clues. 
First, their figures 17 and 18 indicate that the cold infall indeed
tends to be filamentary, especially at high redshifts, while the hot mode is 
more spherical.  They report that the directional signal measuring filamentary
infall in the cold accretion mode is stronger for haloes above $\msh$
while the infall is more isotropic below it.

Second, \citet{keres04} display in their figure 13 the {\it environment} 
dependence of the gas infall modes, showing that the cold and hot modes 
dominate at low and high neighborhood densities respectively.
We learn from this figure that at 
$z<2$ the cold mode dominates for galaxy densities below $n_{\rm gal} \sim 0.3
(\hmpc)^{-3}$ and becomes negligible at larger environment densities, 
while at $z=3$ the cold mode is more pronounced than the hot mode  
for all neighborhood densities up to $n_{\rm gal} \sim 10 (\hmpc)^{-3}$.
The correlation between the environment density and the host halo mass implies
that this $z$ variation of the environment dependence 
could be partly attributed to the finding of a significant residual cold mode
in massive haloes at high $z$ (e.g., their figure 6).    

Third, \citet{keres04} 
show in their figure 16 that the cold accretion is on average 
of {\it higher density\,} than the hot mode.
This is by only a factor $\sim 2$ 
(probably underestimated because they mix small and large haloes),
but since the shock is responsible for a density
increase by a factor of 4, the actual overdensity of the hypothetical
post-shock (or pre-shock) gas is more like $\sim 8$.
Similarly,
\citet{nagai03} find in their simulation of a massive halo that the 
filamentary structure is associated with gas entropy ($\prop T/\rho^{2/3}$)
{\it far below\,} that of the surrounding halo gas. 
%indicating a shorter cooling time in the filaments.
%
In the high-resolution simulation shown in \fig{zinger}, 
we find that the density in the cold streams is actually higher
than the surrounding gas density by two orders of magnitude or more.
The higher gas density in the filaments is associated with a more efficient 
cooling which prevents a shock from forming along the filaments 
(\se{shock_spheri}).

%D NEW filaments and DM. 
We find that the cold gas filaments at the halo outskirts are strongly 
correlated with the DM filaments that feed this halo 
(reported in detail in Seleson \& Dekel, in preparation, hereafter SD05).
These filaments are part of the large-scale cosmic web.
They enter massive halos at high redshift as narrow streams with
a density higher than the halo average by a factor of a few.
The initial overdensity of the gas flowing along the DM filaments
scales with the DM density, while its inflow velocity
is comparable to the halo virial velocity.
As a result, the initial cooling time in the thin filaments is shorter
by a factor of a few than in the surrounding spherical halo, while
the compression time is comparable in the filaments and the host halo.
\Equ{criterion} then implies that the thin filaments have a harder 
time supporting a stable shock. The gas filaments remain cold, and 
become denser as the stream penetrates through the hot medium into the 
halo center. The result is that in massive haloes at high redshift
the critical halo mass for shock heating in the filaments feeding them
is larger than the estimate for a spherical virial shock derived in 
\se{shock_cos}. We provide below a crude estimate for this revised 
critical mass.

%----------------------------------------------
\subsection{Interplay with the Clustering Scale}

\begin{figure}
\vskip 7.2cm
{\includegraphics{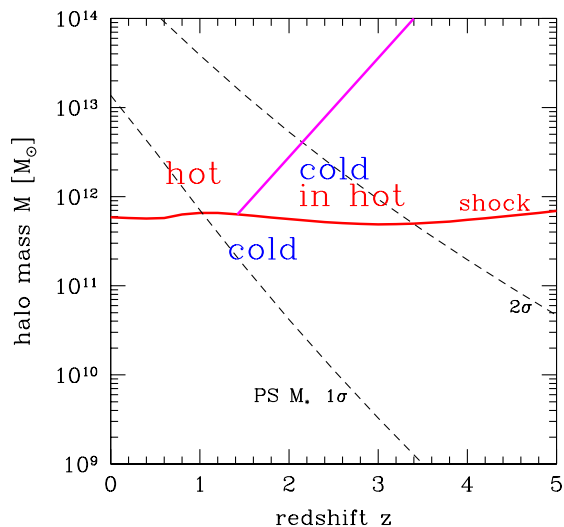}}
\caption{
Cold streams and shock-heated medium as a function of halo mass
and redshift.
The nearly horizontal curve is the typical threshold mass for a stable
shock in the spherical infall from \fig{coolzm}, below which the flows
are predominantly cold and above which a shock-heated medium is present.
The inclined solid curve is the upper limit for cold streams from
\equ{stream} with $f=3$; 
this upper limit is valid at redshifts higher than $z_{\rm crit} \sim 1-2$,
defined by $\msh>f \mps$.
The hot medium in $M>\msh$ haloes at $z>z_{\rm crit}$
hosts cold streams which allow disc growth and star formation, while
haloes of a similar mass at $z<z_{\rm crit}$ are all hot, shutting off
gas supply and star formation.
}
\label{fig:coolzmf}
\end{figure}

What is the reason for the appearance of cold streams in massive haloes
above $\msh$ at high $z$? 
%The cosmological simulations offer a simple explanation, as follows.
% M_*
First recall that there is another characteristic scale 
in the problem --- the scale of {\it nonlinear clustering} $\mps$,
determined by the shape and amplitude of the initial fluctuation power 
spectrum and its growth rate.
The masses for 1-$\sigma$ haloes ($\mps$) and 2-$\sigma$ haloes,
based on \equ{mstar} with $\nu=1$ and $2$ respectively,
are shown again in \fig{coolzmf}.
We note that $\msh \sim \mps$ at $z \leq 1$, while $\msh \gg \mps$ at $z>2$.
This means that $\sim 10^{12}\msun$ haloes are typical at $z<1$ but
they are the highest rare peaks at $z>2$.
We argue that this is responsible for the difference 
in the cold-filament behavior of $M\gsim \msh$ haloes in the two epochs.
Since the large-scale structure of dark matter is roughly self-similar
in time (when measured in terms of $\mps$ and 
the background universal density), we can learn about the
difference between typical and rare haloes by comparing $M\sim \mps$ and
$M \gg \mps$ haloes in a single simulation snapshot. 
One can see in any high-resolution cosmological N-body simulation 
(e.g., the ``Millennium Run", visualized in 
www.mpa-garching.mpg.de/galform/millennium)
that the rare massive haloes tend to be nodes fed by a few intersecting 
relatively narrow filaments which are denser than the virial density 
of these haloes.
On the other hand, a typical $\mps$ halo is commonly embedded in a single
filament of the cosmic web, and this halo thus sees a wide-angle inflow 
pattern in which the matter density is comparable to the virial density
(this is quantified in SD05).
This explains why $\gsim 10^{12}\msun$ haloes are fed by narrow dense
streams at $z>2$ but not at $z<1$.

%7

A crude way to estimate the maximum halo mass for cold streams at a given
redshift is as follows. 
Recall that the critical ratio for shock stability (\equnp{crit}) 
scales roughly like
\be
\frac{\tc}{\tp} \prop \frac{\rho^{-1} T \Lambda^{-1}}{RV^{-1}} \ ,
\label{eq:ratio_crit}
\ee
with $T$, $R$ and $V$ the halo virial quantities and $\rho$ the gas density.
At a given epoch, the typical halo $\rho$ and $R/V$ are roughly independent
of halo mass (based on the definition of the virial radius), 
so with the virial relation $T \prop M^{2/3}$, 
and approximating the cooling function with $\Lambda \prop T^{-1}$
(\equnp{lamfit}), the critical ratio for spherical infall in 
virialized haloes is 
\be
\left(\frac{\tc}{\tp}\right)_{\rm halo} = \left(\frac{M}{\msh}\right)^{4/3} \ .
\label{eq:ratio_halo}
\ee
 
The analogous critical ratio in the dense streams inside a halo of mass $M$,
assuming that $RV^{-1}$ in the streams is the same as in the halo, 
is inversely proportional to the density enhancement 
$\rho_{\rm stream}/\rho_{\rm halo}$ (\equnp{ratio_crit}).
%If the streams' characteristic width is $\prop (f\mps)^{1/3}$ 
%compared to the halo size $\prop M^{1/3}$ (with $f$ a factor of order 
%a few to be determined from simulations),
%the density in the streams feeding a halo of mass $M$ is higher than 
%the halo density by the geometrical factor $(f\mps/M)^{-2/3}$.
Our estimates from N-body simulations indicate that  
$\rho_{\rm stream}/\rho_{\rm vir} \sim (f\mps/M)^{-2/3}$ with $f \sim 3$ 
(SD05).  With \equ{ratio_halo} one obtains
\be
\left(\frac{\tc}{\tp}\right)_{\rm stream} 
=\left(\frac{f\mps}{M}\right)^{2/3} \left(\frac{M}{\msh}\right)^{4/3} \ . 
\ee
For this ratio to equal unity in the streams, the critical halo mass is
\be
\mst \sim \frac{\msh}{f\mps}\msh \ , \quad f\mps<\msh \ .
\label{eq:stream}
\ee
This maximum mass for cold streams is shown in \Fig{coolzmf}.
At low $z$, where $f\mps > \msh$, cold streams exist only for $M<\msh$.
At high $z$, where $f\mps < \msh$, cold streams appear even in $M>\msh$ haloes
where shocks heat part of the gas, as long as $M<\mst$.
The critical redshift $\zc$ separating these two regimes is defined by 
\be
f\mps(\zc)=\msh \ .
\ee

This scenario is consistent with the cosmological hydrodynamical 
simulations.
The shock-heating mass explains the transition from cold to hot at
a given mass roughly independent of $z$, 
and the presence of cold streams above $\msh$ at $z>\zc$ 
explains the dependence of the cold mode on redshift and environment.
Besides its dependence on halo mass, the environment effect 
\citep[e.g.,][]{keres04} may also be due to the survivability 
of cold streams in different environments. 
While streams could survive 
unperturbed in relatively isolated galaxies, they are likely to be 
harassed by the active intergalactic environment in dense groups.
The environment dependence may therefore also reflect variations
in the HOD at a given halo mass.
The properties of cold flows in haloes as a function of halo mass,
redshift, and grouping deserve a detailed analysis
using high-resolution cosmological hydro simulations.

%%%%%%%%%%%%%%%%%%%%%%%%%%%%%%%%%%%%%
%D \section{Feedback and Shock Heating} 
\section{Feedback \& Long-term Shutdown} 
\label{sec:feed}

%D [could be, suggest, likely, can]
Once the halo gas is shock heated in massive haloes, what is the 
process that keeps it hot and maintains the shutdown required by the 
bi-modality?  Is it also responsible for the rise of $M/L$ with mass 
above $\msc$ (and the absence of cooling flows in clusters)?
Several feedback mechanisms can heat the gas. 
We suggest that they have a minimum effectiveness in haloes  
$\sim\msh$. This can be largely induced by the shock heating 
itself, and in turn it can amplify the bi-modality features.
Some of the feedback mechanisms are limited to smaller haloes, while 
others, such as AGN feedback, are likely to be important in more massive 
haloes.  The latter can be triggered by the shock heating and then help 
maitaining the gas hot for a long time.

%---------
\subsection{Below the shock-heating scale}

%D We argue here that the role of the different feedback mechanisms is 
% strongly affected by the shock-heating process. This leads to
% a minimum in feedback effectiveness in haloes of mass comparable 
% to $\msh$, which in turn amplifies the features associated with 
% this scale. AGN feedback may have an especially important role in 
% producing the bi-modality, in concert with the shock heating process.

\no (a) {\bf Supernova feedback}.
Based on the physics of supernova (SN) remnants,
the energy fed to the gas in haloes of $T \sim 10^5$K is proportional 
to the stellar mass despite significant radiative losses \citep{ds86}.
When compared to the energy required for significantly
heating the gas, one obtains a maximum halo virial velocity for 
SN feedback, $V_{\rm SN} \simeq 120\kms$. This is only weakly
dependent on the gas fraction, density or metallicity \citep[][eq. 49]{ds86},
and is therefore insensitive to redshift.
Only in potential wells shallower than $V_{\rm SN}$ can the SN feedback  
significantly suppress further star formation and regulate the process.
\Fig{coolzv} shows $V_{\rm SN}$ and \Fig{coolzm} shows the corresponding 
mass versus redshift.  With an effective $\fb \sim 0.05$, 
the corresponding stellar mass at $z=0$ is $\sim 3.5 \times 10^{10}\msun$, 
practically coinciding with the bi-modality scale. 
The similarity of the SN and shock-heating scales is partly a coincidence, 
because the nuclear origin of the initial SN energy has little 
to do with galactic cooling or dynamics. However, there is an obvious
similarity in the cooling processes and in the asymptotic behavior of a 
SN remnant, which is not a strong function of its initial energy.
%
% FL and origin
The distinct correlations between the properties of galaxies below
$\msc$ indeed point at SN feedback as its primary driver.  These correlations 
define a ``fundamental line",
% over five decades in stellar mass:
$V\propto \Ms^{0.2}$, $Z\propto \Ms^{0.4}$, $\mu \propto \Ms^{0.6}$,
where 
%$V$ is the characteristic internal velocity and 
$\mu$ is surface brightness \citep{kauf03_pop,tremonti04}.
%This is different from the Tully-Fisher relation for
%brighter galaxies \citep{cour05},  
%$V\propto \Ms^{0.3}$, and the lack of systematic
%dependence of $Z$ and $\mu$ on $\Ms$ there.
SN feedback can explain the origin of the fundamental line
%in simple terms 
\citep{dw03} based on (a) the above
energy criterion, which implies $\Ms/M \propto V^2$,
(b) the virial relations (\equnp{virial}),
(c) the instantaneous recycling approximation, $Z \prop \Ms / M_{\rm gas}$,
and (d) angular-momentum conservation, $R_* \propto \lambda R$, with $\lambda$ 
a constant spin parameter \citep*{fe80}.   
%The success of this simple model indicates that SN feedback has an 
%important role below $\msc$.
%associated with certain aspects of the bi-modality, 
%in concert with the shock heating process (\se{bimo}).

\no (b) {\bf UV-on-dust feedback}.   
Also working below $\msh$ are Momentum-driven winds due to radiation
pressure on dust grains, arising from the continuum absorption and 
scattering of UV photons emitted by starbursts or AGNs \citet*{murray04}. 
The dust survives and can provide sufficient optical depth 
if the gas is cold and dense, e.g., in the cold flows below $\msh$, 
which can also provide the starbursts responsible for the required UV flux 
and metals.
% A significant fraction of 
%The cold gas is affected once 
%the luminosity is above an Eddington-like threshold obeying a 
%Faber-Jackson type relation, $L_{\rm max} \prop \fb \sigma^4$.
Since dust grains cannot survive above $\sim 10^6$K,
$\msh$ imposes an upper bound for this feedback.

\no (c) {\bf Photoionization feedback}.
The UV from stars and AGNs ionizes most of the
gas after $z \sim 10$
% \citep{bullock00,loeb01,somerville02,benson03_photo}. 
\citep{bullock00,loeb01}, heats it to $\gsim 10^4$K,
and prevents it from falling into haloes below the Jeans scale
$\Vv \sim 30 \kms$ 
%\citep*{thoul96,quinn96,gnedin00}.
\citep*{gnedin00}.
As the ionization persists for cosmological epochs, the hot gas
evaporates via steady winds from haloes smaller 
than a similar scale \citep{sd03}.
While this is important in dwarf galaxies,
it cannot be very relevant to the bi-modality at $\msh$.

%-----------------
\subsection{Above the shock-heating scale}

\no (d) {\bf AGN feedback}.
The fcat that AGNs exist preferentially in haloes above $\mc$   
may be due to a lower limit for haloes hosting black holes 
\citep*{savvas04}, to starvation of AGNs by SN feedback in haloes 
below $\mc$, or to another reason.
The power emitted from AGNs, e.g., in their radio jets, seems to be
more than necessary for keeping the gas hot.
Given a black-hole mass $M_{\rm BH} \sim 10^7 \msun V_{100}^4$ in a galaxy
of velocity dispersion $V$, and assuming that a fraction $\epsilon$ of this 
mass is radiated away, the ratio of energies is 
$E_{\rm AGN}/E_{\rm gas} \sim 7 \times 10^3\, \epsilon\, 
f_{{\rm b},0.05}^{-1}\, V_{100}^{-1}$. 
For $\epsilon > 10^{-3}$ there seems to be enough AGN energy for affecting 
most of the halo gas. If this energy is released during relatively quiet, 
long phases of self-regulated AGN activity, it can keep the 
gas hot. However, black-hole physics does not seem
to explain the charactersitic scale of bimodality.
Furthermore, had the energy ratio been a measure of feedback 
effectiveness, it would have implied a decline with mass, in conflict with 
the trend of $M/L$. 

The shock heating of the gas into a dilute medium
is likeley to make it vulnerable to heating and pushing by the central 
energy source, thus providing the {\it trigger\,} for effective AGN 
feedback.  Simulations of winds in a two-phase
medium demonstrate that the dilute phase is pushed away while the 
dense clouds are hardly affected \citep{slyz05}. 
This behaviour is likely to be generic, though
the mechanism by which the energy released near the black hole
is spread in the halo gas is an open issue 
%\citep*{ruszkowski04_agn,begelman04,scannapieco04}.
\citep*{ruszkowski04_agn,scannapieco04}.
If so, the feedback efficiency may be driven by the relative fraction of 
hot gas rather than the actual AGN energy.
Figure 6 of \citet{keres04} shows that the 
hot fraction varies roughly $\propto M^{1/2}$, implying  
$M/L \prop M^{1/2}$ near and above $\msh$, in qualitative agreement with
the observed trend (\se{intro}, item k).
In this scenario, 
%the coupling of AGN feedback with the hot gas is responsible
% for the desired shutdown, and 
the shutdown scale arises naturally from the shock heating.

\no (e) {\bf Two-phase Medium}.
Given that the cooling function peaks near $\gsim 10^4$K,
the virialized gas at $\gsim 10^6$K develops a two-phase medium, with cold, 
dense clouds pressure confined within the hot, dilute medium 
%\citep{field65,fall85,murray90}.
\citep{field65,fall85}.
The cloud sizes and evolution are affected by thermal conductivity and
dynamical processes \citep{voigt04_cond}.  
This 
%phenomenon acts like ``feedback" in the sense that it
can help 
explaining the bright-end truncation of the blue sequence \citep{maller04}.
Some of the gas is locked in the orbiting
clouds and the density $\rho_{\rm hot}$ of the hot gas is reduced,
slowing the cooling and the infall.
Approximating \equ{crit} with $\tc/\tp \prop \rho^{-1} \Tv^2$, and recalling
that $\Tv \prop \Mv^{2/3}$, the longer cooling time makes the
critical mass for further shock heating ($\tc/\tp\sim 1$)
smaller by a similar factor, $\msh \prop \rho_{\rm hot}^{3/4}$,
namely it enhances the shock stability.
%(same as $\msh \prop \fb^{3/4}$ in \equ{mapp}).
The gas may be kept hot over longer periods
by repeating shocks due to continuous accretion into the halo,
which may alleviate the need for AGN feedback.
Still, a necessary condition for hot gas 
is the initial shock heating, i.e., being in a halo above $\msh$.

%8
\begin{figure}
\vskip 5.8cm
{\includegraphics{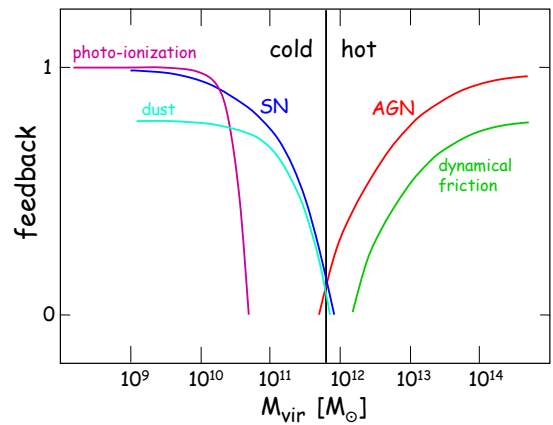}}
\caption{
The ``strength" of the various feedback processes at $z=0$, e.g. 
referring to the
fraction of the initial gas that has been heated or removed (schematic).
Different feedback processes are effective below and above the critical
shock-heating scale $\lsim 10^{12}\msun$, and the feedback efficiency 
is at a minimum near this scale, giving rise to a minimum in $M/L$ there.
}
\label{fig:feed}
\end{figure}

\no (f) {\bf Dynamical-friction feedback}.
Another heating source is the dynamical friction acting 
on galaxies in a halo core.  The energy transferred is
comparable to that required for preventing cooling flows in cluster centres
\citep*{elzant+kam04}.
The gas response to dynamical friction, unlike the DM response,
has a sharp peak near a Mach number of unity \citep[][Fig.~3]{ostriker99},
namely when the gas is heated to near the virial temperature in
$M>\msh$ haloes and not in smaller haloes hosting cooler gas.
As groups occur above a critical halo mass that roughly coincides
with $\msh$ at $z \leq 1$, this feedback appears
almost simultaneously with the hot medium, which
then serves as the vulnerable victim of the same feedback 
(as for AGN feedback).

% schematic figure of feedback 
\Fig{feed} is an illustration of the strength of the different feedback 
processes, referring to the gas fraction that could have been heated  
at $z\sim 0$.  The figure highlights the fact that different 
processes dominate below and above $\msh \lsim 10^{12}\msun$. 
The transition from cold to hot infall has a crucial role in 
determining the feedback efficiencies near the critical mass;
it induces a minimum in feedback efficiency 
at a critical scale $\mfb \sim \msh$ and drives the shapes of the curves
about this minimum.
At higher redshifts this minimum becomes wider and deeper
but centered on a similar critical mass.
%D We address the effects of feedback on the bi-modality features
% in \se{bimo}.

%%%%%%%%%%%%%%%%%%%%%%%%%%%%%%%%%%%%%%%%%
\section{The Origin of Bi-Modality}
\label{sec:bimo}

\subsection{A scenario from the assumed ingredients}

We propose that the cold flows and shock heating
play a key role in producing the observed bi-modality features.
These features are emphasized by the similarity between
the scales associated with shock heating, feedback and clustering.
Based on our current understanding of these physical processes, we 
assume the validity of the following:

\no (a) {\bf A new mode of star formation}.
The collisions of the (partly clumpy) cold streams with each other 
and with the inner disc are assumed to produce starbursts, analogous to  
collisions of cold gaseous discs or clouds.  These collisions are 
expected to produce isothermal shocks,
behind which the rapid cooling generates dense, cold slabs 
where the Jeans mass is small.
While the details are yet to be worked out, we assume
that such a mode of star formation may be responsible for much of the
stars in the universe. In some cases, this can be an enhanced quiescent mode,
leaving the disc intact without producing a big spheroid, an in other
cases it may resemble the starbursts associated with mergers.

\no (b) {\bf Hot forever}.
Once the gas in a massive halo is shock heated to near the virial temperature, 
it is assumed to be hot forever.
This is based on the slow cooling time of the dilute hot medium
and its vulnerability to AGN feedback, while cold, dense clouds and streams
could be better shielded against winds and ionizing radiation.
The shock heating is thus assumed to {\it trigger\,} a shutdown of  
all modes of star formation in haloes where cold streams 
do not prevail. 

\no (c) {\bf Cold streams in a hot medium}.
Cold streams in haloes above $\msh$ (\se{simu}) are assumed 
to supply cold gas for further disc growth and star formation,
preferentially before $\zc \sim 2$ and in isolated galaxies.
After $\zc$, especially in groups, cold streams are
suppressed and a complete shutdown of star formation is assumed to follow.

%9
\begin{figure}
\vskip 11.9cm
\includegraphics{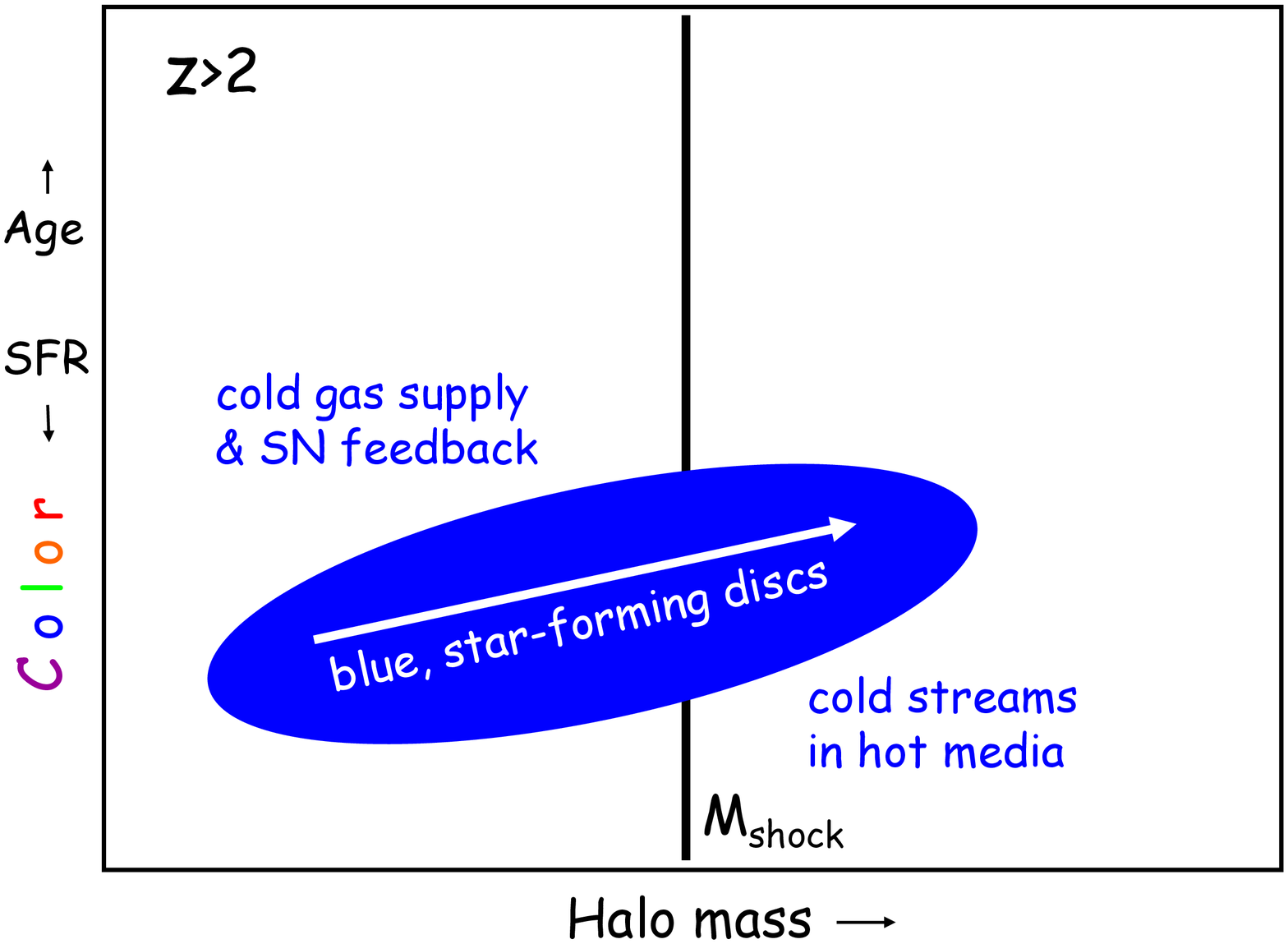}
\includegraphics{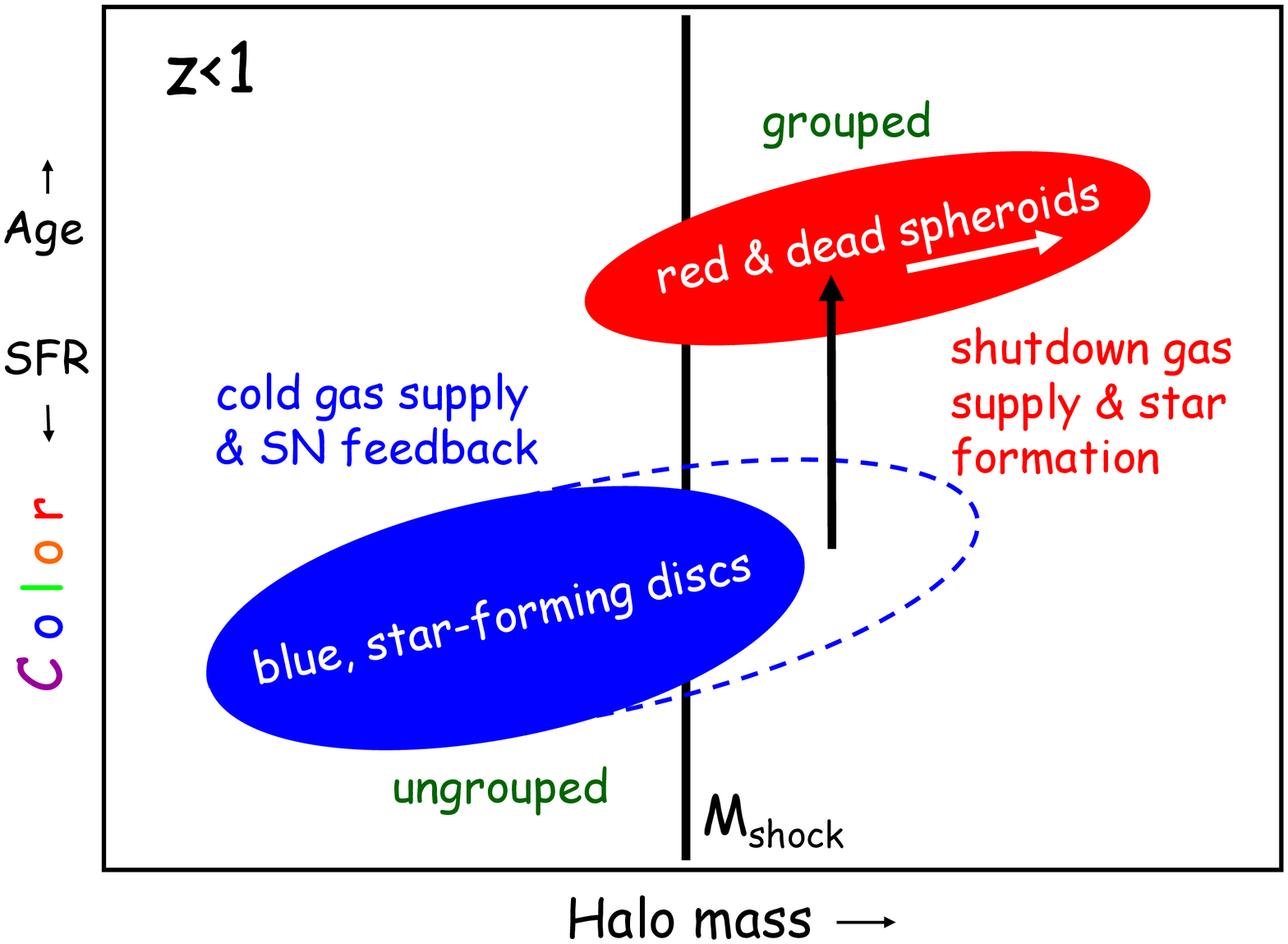}
\includegraphics{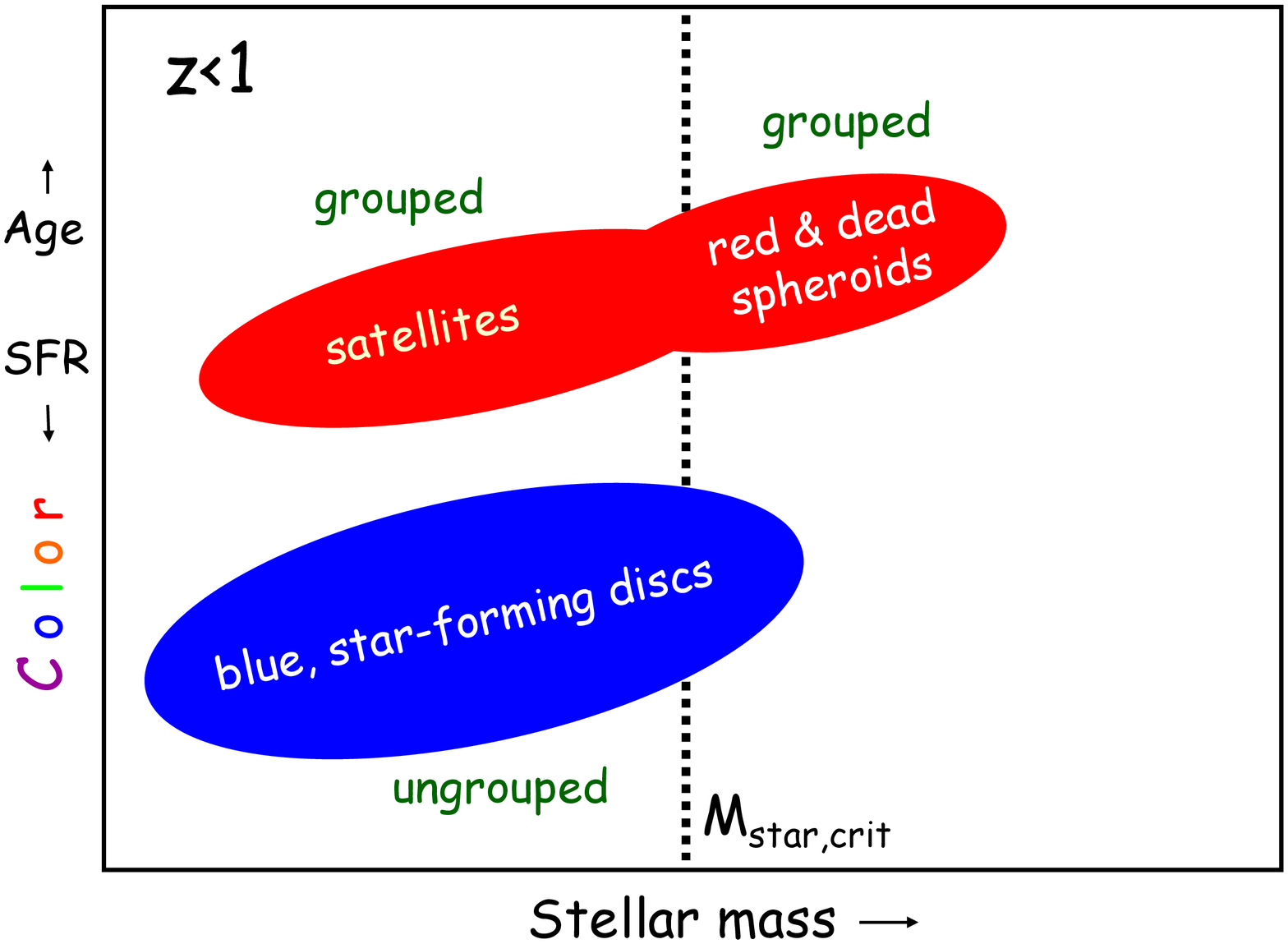}
\caption{
Schematic illustration of the origin of the bi-modality in color (or SFR
or stellar age) as a function of halo mass.
At $z>2$ (top), continuous gas supply, regulated by supernova feedback,
makes the galaxies evolve along the blue sequence, which extends 
beyond the shock-heating mass due to cold streams in hot media.
At $z<1$ (middle), in the absence of cold streams above $\msh$, 
the shock-heated gas is kept
hot by AGN feedback, gas supply and star formation shut down, and the 
stellar population passively turns red \& dead. Gas-poor mergers stretch
the red sequence toward larger masses. 
When the halo mass is replaced by stellar mass (bottom), 
the red sequence is stretched toward small stellar masses 
due to satellite galaxies sharing a common halo. 
The color is correlated with the environment density via the halo mass,
with the minimum group mass being comparable to $\msh$.
}
\label{fig:bimod}
\end{figure}

% origin of bi-modality
The {\it bi-modality\,} in color (or stellar age, or SFR) versus mass, 
the correlation with the environment and the evolution with redshift, 
all emerge naturally from the early efficient star formation followed
after $z\sim 2$ by the {\it abrupt shutdown\,} in haloes above $\msh$,
which typiclaly host groups. This is illustrated in \fig{bimod}:  

\no (a) {\bf The blue sequence\,}
is dominated by galaxies in haloes below $\msh$, as they grow 
by accretion/mergers.  Cold flows lead to early disc growth and star formation,
which is regulated by SN feedback over cosmological times.
Galaxies can get very blue because of repeating starbursts
due to the clumpy gas supply 
and the interplay between infall, starburst and outflow.
 
\no (b) {\bf Bright blue extension}.
Some galaxies continue to be fed by cold streams even when they are
more massive than $\msh$, extending the blue sequence beyond $L_*$. 
This occurs especially at $z > \zc \sim 2$, when the streams feeding 
high-$\sigma$ haloes
are relatively narrow and dense, resulting in massive starbursts. 

\no (c) {\bf The red sequence}.
Once a halo is more massive than $\msh$, halo gas is shock heated; 
it becomes dilute and vulnerable to AGN feedback. 
At $z < \zc$, where cold streams are suppressed,
gas supply from the host halo shuts off, preventing any further growth
of discs and star formation. 
If residual cold gas has been consumed in earlier mergers,
there is a total shutdown of all modes of star formation above $\msh$,
allowing the stellar population to passively turns ``red and dead" into the red
sequence. 
The massive tip of the blue sequence at $z > \zc$ becomes the massive
tip of the red sequence at $z < \zc$, explaining the bright-end truncation
of the blue sequence at low $z$, and the apearance of brighter,
very red galaxies already at $z\sim 1$.
Subsequent growth along the red sequence is induced by gas-poor mergers
uncontaminated by new blue stars.

\no (d) {\bf Faint red extension}.
When color is plotted against stellar mass, the bi-modality 
extends to smaller galaxies which are typically satellites of the
central galaxies in common haloes. In haloes below $\msh$, accretion
onto satellite galaxies can keep them on the blue sequence for a while.
In haloes above $\msh$, where gas supply stops and  
the environment density is high, the satellites too become red \& dead.

\begin{figure}
\vskip 11.8cm
\includegraphics{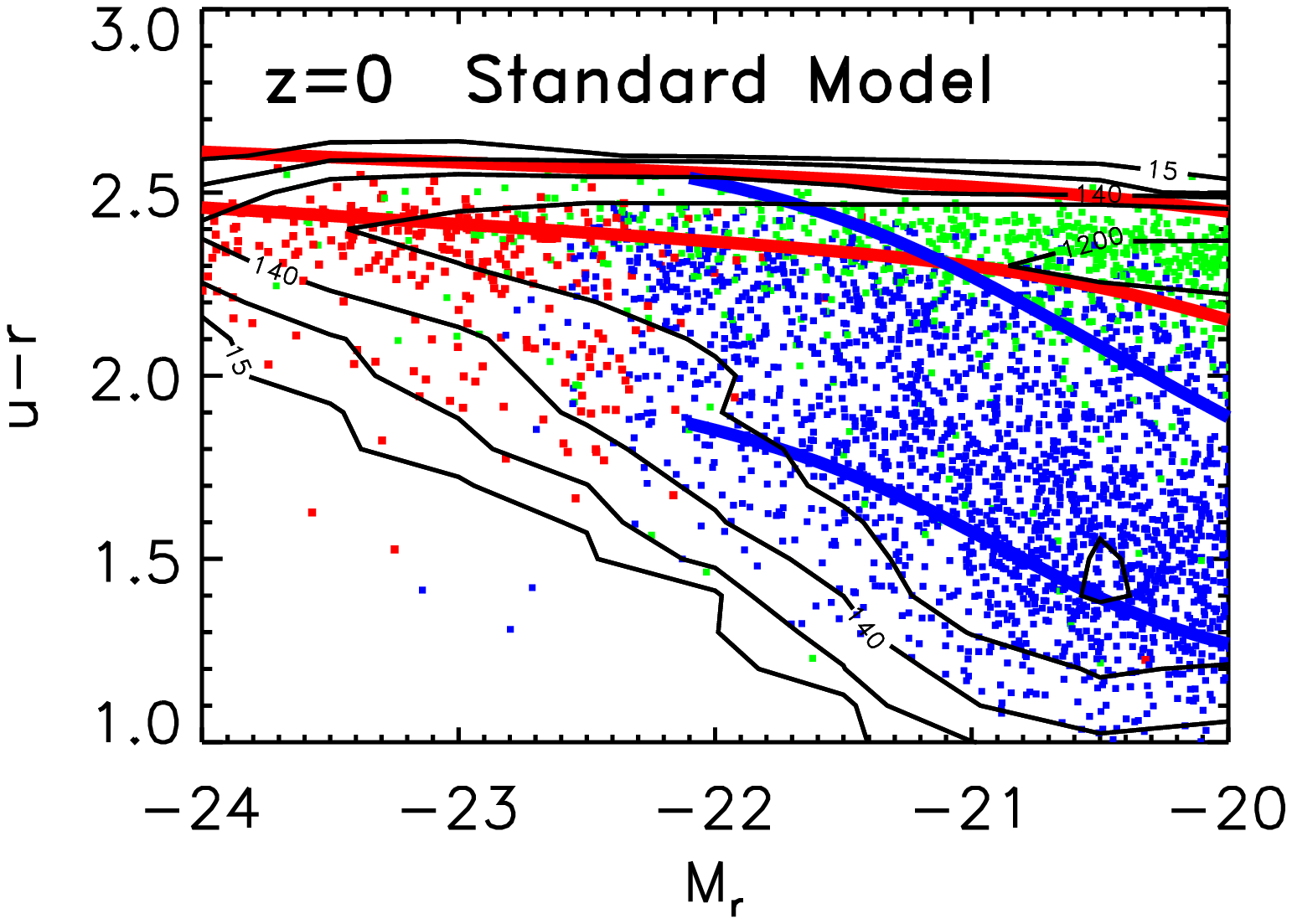}
\includegraphics{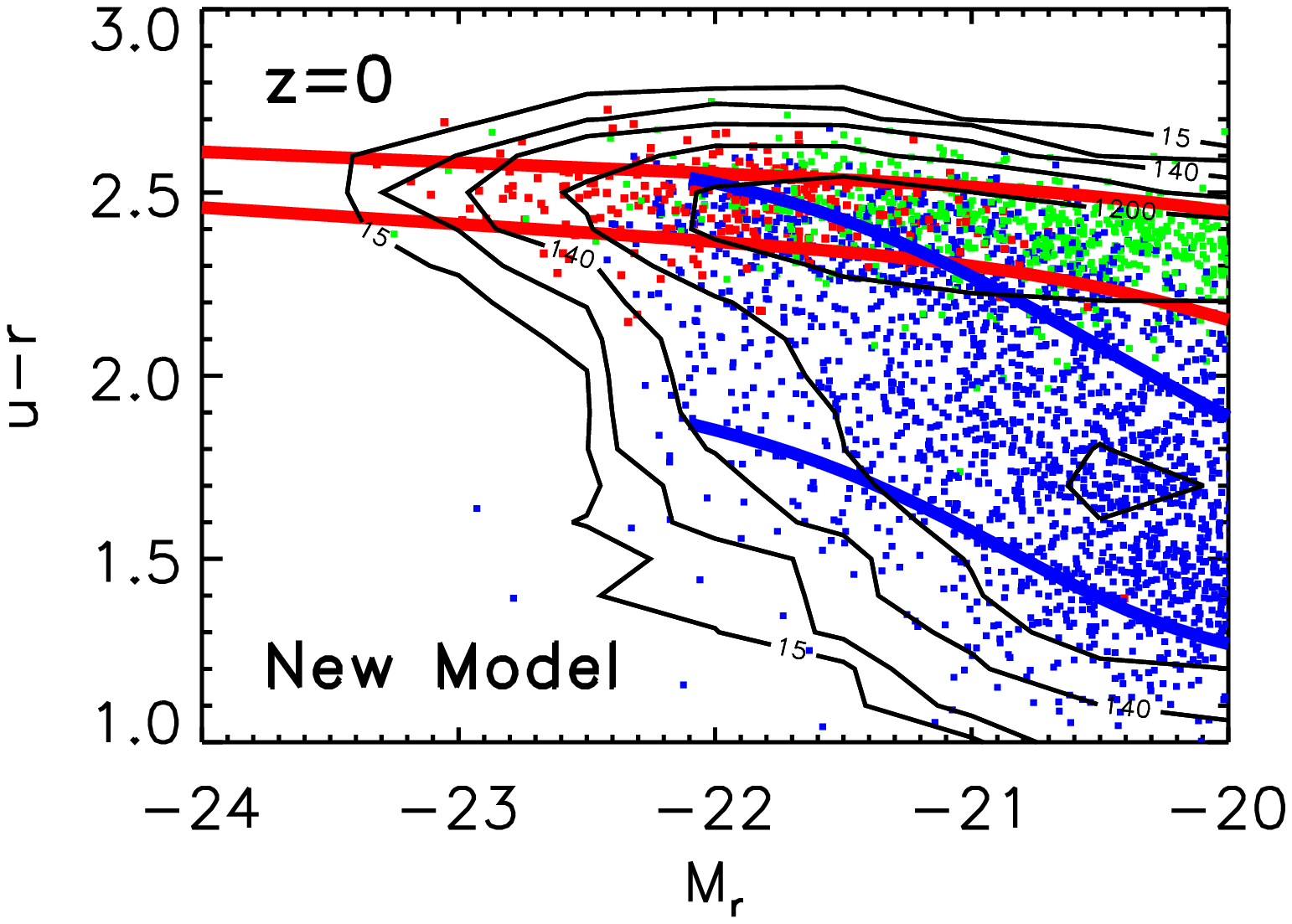}
\includegraphics{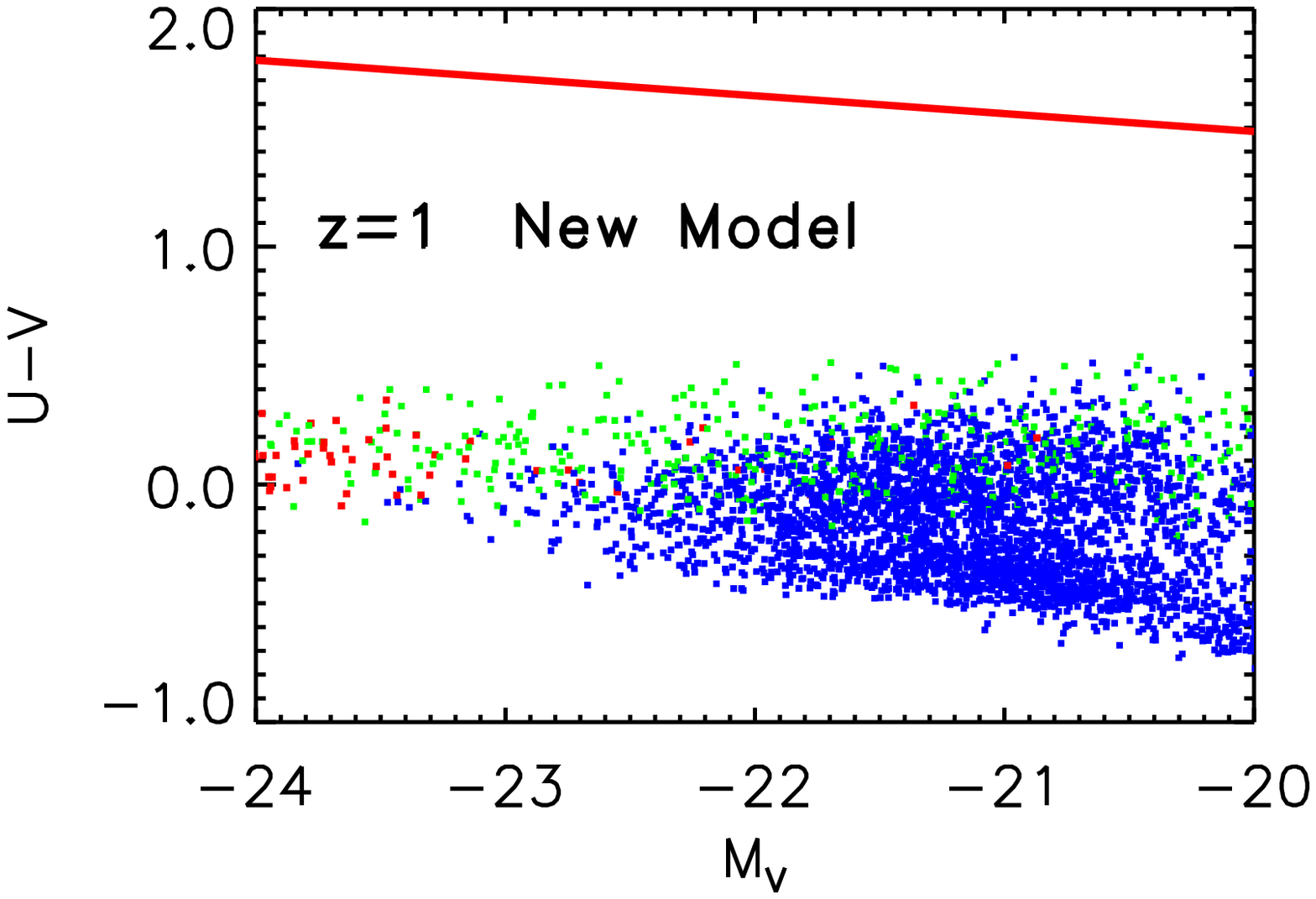}
\caption{
Color-magnitude diagrams from a hybrid semi-analytic/N-body simulation
\citep{cattaneo05}
demonstrating the success of the proposed scenario in reproducing 
the main bi-modality features along the lines of \fig{bimod}. 
{\bf Top:} Based on the ``standard" version of the GalICS SAM.
{\bf Middle:} The result of incorporating a shutdown in star 
formation above $\msh \sim 10^{12}\msun$ after $\zc \sim 3$, and allowing 
for star formation even in $>\msh$ halos prior to $\zc$.
{\bf Bottom:} Same at $z=3$. 
Blue dots refer to galaxies in haloes below $\msh$, while 
red and green dots are galaxies in halos above $\msh$, central and
satellite galaxies respectively.
The halo mass correlates with number density in the environment.
Contours mark number density of galaxies in the diagram.
The color curves mark the main bodies of the blue and red sequences in the 
SDSS data \citep{baldry04}.
}
\label{fig:cattaneo}
\end{figure}

% C05
In an associated paper \citep[][C05]{cattaneo05}
we have implemented the proposed new physics 
in a hybrid SAM/N-body simulation \citep[GalICS][]{hatton03}.
The main new ingredient is a shutdown in gas cooling and star
formation above $\msh \sim 10^{12}\msun$ after $\zc \sim 3$, while allowing for
efficient star formation by cold streams in more massive halos prior to $\zc$.
The revision yields excellent fits to the observed features at low and high
redshifts (C05).
\Fig{cattaneo} shows one example of the results ---  color-magnitude diagrams
which demonstrate the success of this model along the lines envisioned in
\fig{bimod}.
The top panel highlights the main deficiencies of the ``standard" model at
$z=0$: an excess of bright blue galaxies accompanied by a shortage of 
red-enough galaxies compared to the SDSS data \citep{baldry04}.
The ``new" model puts the blue and red sequences where they should be
at $z=0$, with a proper truncation at the bright-blue end and appropriately
red colors in the red-sequence.
The $z=3$ diagram shows the predicted bright blue galaxies (which were absent
in the ``standard" model, C05).
The colors distinguish between galaxies in haloes below (blue) and above
$2\times 10^{12}\msun$, comparable to the mass separating haloes hosting
field galaxies and groups.
This indicates that the model recovers the correlation between color
and environment density.
It also recovers the bi-modality in bulge-to-disc ratio (C05).

\subsection{Origin of Observed Features}

% gap 
The {\it color-magnitude\,} bi-modality emerges from the color-mass 
bi-modality lying at the basis of the model.
The red satellites extend the luminosity range over which the blue and red 
sequences co-exist and highlight the {\it gap\,} in color.
This gap is amplified because the galaxies making the transition into the 
red sequence once their haloes become $>\msh$ tend to be the merger remnants 
with big bulges and AGNs from the {\it red\,} tip of the blue sequence. 

% bimodality in B/D
The correlated bi-modality in {\it bulge-to-disc ratio} can be understood
in the scenario where most of the big spheroids in the red sequence are 
passively aged galaxies that have grown massive stellar components
already in the blue sequence (C05).
The transition to the red sequence is likely to be made by galaxies with 
big spheroids because 
(a) they have consumed their gas in the same mergers that produced the 
spheroids,
(b) these mergers tend to occur in big haloes hosting groups 
where shock heating stops the gas supply, and (c)  
these spheroids contain massive black holes that can keep the gas hot. 

%SFR and environment    
The strong anti-correlation of {\it star-formation rate\,}
(and blue color) with the number density in the 
{\it environment\,} is a natural outcome of the mass dependence. 
The key is that the minimum halo mass for groups at $\sim 10^{12}\msun$ 
is comparable to $\msh$ at $z \leq 1$.
The strong dependence of cold gas supply on host-halo mass 
can therefore be responsible for the distinction between the SFR 
in field and clustered galaxies 
(which is thus predicted to be limited to late times).
The galaxies dominating low-HOD haloes below $\msh$ 
enjoy cold gas supply leading to discs forming stars. 
The galaxies populating groups of subhaloes,  
typically in haloes above $\msh$ at $z \leq 1$,
suffered starvation of external gas supply and lost their internal
gas in mergers, thus stopped forming stars and passively evolved
to the red sequence. 
The faint end of the red sequence, which is preferentially present 
in high environment densities, is due to the starvation of satellite 
galaxies in the high-HOD haloes above $\msh$. The model naturally
predicts the seondary bi-modlaity seen along the red sequence
\citep[along the lines of][]{berlind04}.

%B/D and environment     
The classic {\it morphology-environment\,} relation,
traditionally attributed to the correlation of merger rate with the
environment, may also be viewed as a result of the cold-flow phenomenon.
New discs are predicted to form only in haloes below $\msh$, namely in 
``field" galaxies, and not in the group haloes above $\msh$.
On the other hand, the frequent major mergers in groups
help build the big spheroids preferentially there.
The abrupt shutdown of star formation above $\msh$ makes
the color-environment correlation stronger than the
morphology-environment correlation.

% clustering
The environment dependence highlights an interesting cross-talk between the
{\it clustering\,} and gas processes.
The distinction between haloes hosting a single dominant galaxy and
haloes hosting groups has traditionally been attributed
to gas cooling on a dynamical time scale \citep{ro77}.
Our shock-stability analysis helps quantifying this idea.
However, it seems from N-body simulations that the gravitational DM
clustering process has a parallel role \citep{krav04_hod}.
The HOD of subhaloes develops a transition from single to multiple occupancy
near a comparable halo mass, associated with the current $\mps$.
The HOD of subhaloes is similar to the HOD of galaxies deduced from the
observed correlation function for galaxies
(\se{intro}, item $i$). One can conclude that in haloes above 
$\sim 10^{12}\msun$
the potential wells associated with the DM subhaloes provide
the sites for the fragmented gas collapse on the scales preferred
by cooling, thus emphasizing this scale as the minimum scale for groups.
The coincidence between these scales is behind the environment dependence 
of the bi-modality features.
We have discussed above the other possible role of $\mps$ in the appearance
of cold filaments in hot haloes at high $z$ (\se{simu}) and in some of the
feedback processes (\se{feed}).

%M/L
The minimum in mean halo {\it mass-to-light\,} ratio $M/L$ near
$\sim 10^{11-12}\msun$ 
can be attributed to the maximum in gas supply near $\msh$
dictated by the shock heating above this mass and the associated
minimum in feedback there (\fig{feed}).
While SN feedback gets stronger toward smaller haloes, 
AGN feedback pumps energy more effectively into the shock-heated medium in
more massive haloes.
The ``fundamental line" due to SN feedback below $\msh$
corresponds to $M/L \propto M^{-2/3}$ \citep{dw03},
while the transition from cold to hot infall indicates $M/L \prop M^{1/2}$
above $\msh$, in the ballpark of the findings from 2dF (\se{intro}, item k).
This settles the discrepancy between the halo mass function and the 
galaxy luminosity function both below and above the bi-modality scale.
Our model predicts a similar behavior of $M/L(M)$ at $z \sim 1$,
to be revealed by spectroscopic surveys
\citep[such as the DEEP2/3][]{madgwick03,coil04},
which have already confirmed the predicted invariance of the 
bi-modality mass in this redshift range (\fig{coolzmf}).

% SN vs shock -- effect on fb at high z
The massive starbursts at high $z$, primarily due to
cold streams, are helped by an increase in $\msh$ itself.
At high $z$, the upper bound for SN feedback becomes $\ll \msh$ (\fig{coolzm}), 
allowing for a more efficient gas accretion at $\lsim \msh$.
If $\fb$ is $0.13$ instead of $0.05$ at $z\sim 3$,
the value of $\msh$ doubles (\equ{mapp}).
Another factor proportional to $\fb$ enters  
when translating from halo to stellar mass, yielding a total
increase of $\sim 5$ in the critical stellar mass at high $z$.
We thus expect a strong star-formation activity at high $z$ in galaxies 
with stellar masses exceeding $\sim 10^{11}\msun$.

%time dep Madau
The global {\it star formation history} could be derived from
the predicted SFR as a function of mass and redshift, 
convolved with the time evolution of the halo mass function in the 
given cosmology.
With the prediction that haloes $\lsim \msh$ are
the most efficient star formers, and with the Press-Schechter estimate that
haloes of such a mass typically form at $z\sim 1$ (\fig{coolzm}),
the star formation density is predicted to peak near $z\sim 1$,
with a relatively flat behavior toward higher
redshifts and a sharp drop toward lower redshifts, as observed.
In particular, the cumulative stellar density seems to stop growing quite 
abruptly at $z\sim 1$ \citep{dickinson03_sfr}, when the typical forming haloes
become larger than $\msh$ and the cold flows are suppressed.
The growth of the hot fraction as a function of halo mass \citep{keres04}
can be translated to the drop in star-formation rate after $z\sim 1$.
The ``downsizing" of star formation in galaxies is also helped by the shutdown
of star formation above $\msh$ while smaller galaxies 
can make stars also later, until they fall into bigger haloes.

%%%%%%%%%%%%%%%%%%%%%%%%%%%%%%%%%%%%%
\section{Discussion: other implications}
\label{sec:discussion}

Possible implications on open issues in galaxy formation where further 
study is desired are worth mentioning. 

\no{\bf X-rays}.
The shock-heating scale may be detectable in soft X-rays,
as the lower limit for galaxies and groups containing hot halo gas. 
In turn, the suppression of heating below $\msh$ 
explains the missing soft X-ray background
\citep{pen99,benson00}.
We predict a noticeable suppression of X-ray emission in the range 
$5\times 10^5$ to $2\times 10^6$K.

\no{\bf Lyman-alpha emission}.
The cold ($\sim 10^4$K) flows may instead be an
efficient source of \lya radiation, possibly associated with observed
\lya {\it emitters\,} at high redshift
\citep{kurk03_lya,nilsson05}. %D7
It has been argued based on SPH simulations \citep{fardal01,furlanetto03}
that the flows radiate their infall energy mostly in \lya before they blend
smoothly into the discs.
Our Eulerian simulations (ZBDK05) indicate that the streams do eventually 
shock in the inner halo. This produces X-ray, but, given the high density 
there, the X-ray radiation is likely to be confined to an ionized
Str\"omgren sphere of a few kiloparsecs.
This energy eventually transforms into \lya radiation, which could
propagate out via thermal broadening and systematic redshifts.
A study involving radiative transfer is required.

\no{\bf Damped Lyman-alpha systems}.
The possible association of massive cold flows with damped Ly-$\alpha$ 
systems \citep{prochaska03} should be addressed in cosmological simulations.

\no{\bf LIRGS}.
Cold flows may explain the massive starbursts associated
with LIRGs at $z \lsim 1$ \citep{hammer04}.
If half the stars in today's discs were formed in such
LIRGs, and the majority of galaxies today are fragile discs,
than many of the LIRGs could not have been produced by violent
major mergers. The cold streams may provide a less violent
starburst mechanism not associated with the destruction of discs.
Simulations that properly incorporate cold streams should 
be confronted with these data.

\no{\bf Angular momentum}.
The proposed scenario may set the stage for solving the angular-momentum 
puzzle --- the over-production of low angular momentum spheroids in 
current cosmological simulations \citep{navarro00_am}.
The solution should involve the removal of baryons with low angular momentum.
In small galaxies, SN feedback can blow the gas away from their small
building blocks, which are otherwise the main source of low angular
momentum via many minor mergers \citep*{mds02,md02}.
In galaxies near $\msh$, we find from cosmological simulations 
(ZBDK05, see \fig{zinger}) that the low angular momentum gas is typically 
associated with the shock-heated medium, which can be prevented
from cooling by AGN feedback. The cold streams come
from larger distances with $\sim 50\%$ higher specific
angular momentum, appropriate for producing extended discs. 
The feedback effects, both below and above the critical scale,
have not been properly simulated yet because of incomplete treatment of the 
micro-physics.

\no{\bf Cold clouds}.
The formation of discs by a clumpy cold gas phase may have the following 
implications.
(a) It may explain the starbursts responsible for very-blue galaxies.
(b) It may help explaining the bright-end truncation of the luminosity 
function (\se{feed}).
(c) The dynamical friction bringing the clouds into the disc 
transfers energy into the halo, which may help explaining the  
discrepancy between the steep cusps predicted by N-body 
simulations and the flat cores indicated by rotation curves
in low-surface-brightness galaxies \citep*{ddh03,dadb03,elzant04,ma04}.
(d) The same process may lower the predicted maximum rotation velocity
in discs at a given luminosity, balancing the adiabatic contraction
of the dark halo, and thus repair the zero-point offset in models of
the Tully-Fisher relation \citep[e.g.][]{klypin02,abadi03_tf,dutton05}.
(e) This may explain the lack of anti-correlation between
the residuals in velocity and radius at a given luminosity \citep{cour99},
indicating comparable contributions of disk and dark halo to the 
gravitational potential at the effective disc radius \citep{dutton06}.

\no{\bf Dust lane}.
Edge-on discs above $\Vv \simeq 120\kms$ show a well-defined dust lane,
while less massive discs show diffuse dust above and below the disc 
\citep*{dalcanton04}.
A thick, turbulent, dusty gas phase is indeed expected
when SN feedback is effective, and when cold streams shock 
and produce stars, both predicted below an appropriate scale.

\begin{figure}
\vskip 8.4cm
{\includegraphics{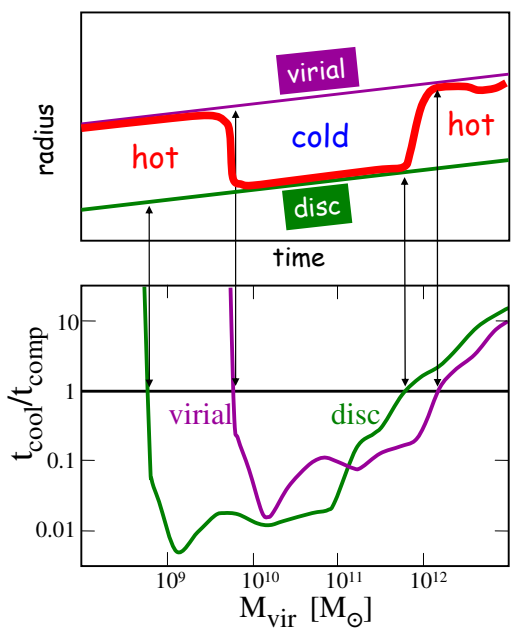}}
\caption{
Shock stability as a function of mass at $z=0$.
{\bf Bottom:} The ratio of rates, $\tc/\tp$,
as a function of halo mass derived at the disc ($Z=0.1$,
$\us=0$) and at the virial radius ($Z=0.03$, $\us=0.15$).
The cooling rate is assumed to vanish for $T<10^4$K.
A stable shock is possible once $\tc/\tp >1$.
{\bf Top:} An illustration of the evolution of shock radius 
between the disc vicinity and the virial radius
as the halo mass grows in time, based on the stability criterion shown
in the bottom panel.
}
\label{fig:coolm}
\end{figure}

\no{\bf Shock heating in dwarf haloes}.
The cold/hot infall and feedback processes are expected to 
give rise to two scales characterizing dwarf galaxies.
The lower bound at $\Vv \sim 10-15\kms$ \citep[e.g.,][Fig.~3]{dw03}
is commonly attributed to the drop in atomic cooling
rate below $\sim 10^4$K. We propose that this involves 
shock heating, in analogy to $\msh$ discussed above.
\Fig{coolm} (bottom) shows the quantity relevant to shock stability,
$\tc/\tp$ [\equ{crit}], 
as a function of halo mass, now stretched to low masses.
The molecular-hydrogen cooling rate, relevant below $10^4$K,
is weaker and may actually be eliminated after $z\sim 10$
due to molecule dissociation by the UV background \citep{haiman96}.
The stability is evaluated at $z=0$ both in the disc vicinity ($r=0.1\Rv$,
$Z=0.1$, $\ust=0$, and near the virial radius ($Z=0.03$, $\ust=1/7$).

%10

% evolution of shock
The top panel illustrates the deduced evolution of shock radius
between the disc and the virial radius as the halo mass grows
in time.  Haloes below $\sim 10^9\msun$ have stable virial shocks.
As the halo grows to $\sim 6\times 10^8\msun$,
the conditions near the disc become unfavorable for a
stable shock, but the virial shock persists.
Only when the halo becomes $\sim 6\times 10^{9}\msun$ 
shock heating stops completely. 
Thus, the shock heating prevents star formation in smaller haloes even 
before the epoch of re-ionization.
Once the first stars form in the dense centres of big clouds, their UV 
radiation reionizes the gas, dissociates molecules, and helps preventing 
cooling.  This provides a natural explanation for the absence of luminous
galaxies with haloes below $\sim 10^9\msun$, predicting a large population 
of mini haloes with $\Vv < 10\kms$ which are completely dark.
There is only a narrow window, $6\times 10^9 - 6\times 10^{11}\msun$,
for haloes that allow cold flows and can form luminous discs at low redshifts.
This range is somewhat broader at high redshifts, with the lower bound 
dropping below $10^9\msun$ and the upper bound rising above $10^{12}\msun$ 
prior to $z\sim 3$. 
% DDH
Within this window, it seems that below $\Vv \sim 30\kms$ most of the 
haloes are totally dark\footnote{This can be deduced from the discrepancy 
between the flat faint-end luminosity function and the steep halo
mass function predicted in $\Lambda$CDM, given the Tully-Fisher like
velocity-luminosity relation of dwarf galaxies \citep[e.g.,][Fig.~3]{dw03}.}
and the others are populated by gas-poor dwarf spheroidals.
This is the scale predicted by photoionization feedback (\se{feed}).

%%%%%%%%%%%%%%%%%%%
\section{Conclusion}
\label{sec:conc}

%------------------------------
\subsection{Summary of results}

% summary: shock heating
The classic argument of cooling on a dynamical time scale \citep{ro77,wr78}, 
with order-of-magnitude estimates of the time scales involved, 
provided an inspiring qualitative upper bound for luminous galaxies, 
at a halo mass of $M \sim 10^{12-13}\msun$.
An analytic study of the actual shock-heating process 
\citep[][and this paper]{bd03}
now yields a more concrete halo critical scale at
$M \simeq 6\times 10^{11}\msun$, somewhat
smaller than the original estimate.
The criterion for critical shock stability, 
\be
\tc^{-1} = \tp^{-1} \ ,
\ee
is a balance between the cooling rate and the post-shock compression rate,
which restores the pressure supporting the shock against gravitational 
collapse.
The compression time is somewhat larger than the crossing time
at the shock position. 
The absolute magnitudes of these time scales are irrelevant -- they
could in principle both be longer than the Hubble time, because what 
matters for shock heating versus cold flows is only the relative rates 
of the competing processes.
The most relevant critical scale is obtained in the inner halo,
because as the halo grows, the shock first becomes stable in the
inner halo, and it then propagates outward to the virial radius.
Haloes of mass below the threshold mass build discs in their centres by cold
flows, while in haloes above the threshold much of the gas is shock-heated.
These results are confirmed by spherical hydrodynamical simulations. 
The same phenomenon is seen at a comparable scale
in general cosmological hydrodynamical simulations.
They reveal that in haloes near the critical scale, and even in larger
haloes preferentially at $z\geq 2$ and especially in field galaxies, 
cold streams along the filaments feeding the galaxy penetrate  
through the hot medium, and allow further disc growth and star formation. 
 
% scenario
The interplay between these cold flows and shock-heating, 
the gravitational clustering scale, and the different feedback processes 
acting below and above a similar mass scale,
is proposed to be responsible for the robust bi-modality 
imprinted on the observed galaxy properties.
Cold flows in haloes much bigger than the clustering scale
allow massive starbursts at $z\geq 2$, while shock heating in
comparable haloes at later times shuts off star formation and leads
to big red galaxies.
While supernova and radiative feedbacks regulate star formation below 
the critical scale, the presence of dilute, shock-heated gas in more 
massive haloes allows the AGN feedback (or another energetic source 
capable of affecting big galaxies) to keep the shock-heated gas hot
and prevent further disc growth and star formation.
The observed bi-modality and many of the related phenomena are argued to 
arise naturally from such a scenario (\se{bimo}).
% small masses
The shock-heating process also plays a role in introducing a lower
bound to haloes hosting galaxies, at $\sim 10^9\msun$. 
The mass range where disc galaxies can form today turns out
to be quite narrow, between a few times $10^9\msun$ to slightly below 
$10^{12}\msun$.

%--------------------
\subsection{Re-engineering of SAMs}

%SAM
Once the new physical processes 
are properly incorporated in the detailed models of galaxy formation,
they solve many of
the apparent conflicts between theory and observation.
At a crude level, one might have naively thought that since the cooling
time is anyway assumed to be shorter in smaller haloes, 
the details of the cold-flows and shock heating 
would not matter much to the final result. 
However, a closer inspection reveals that
there are several key features which make a qualitative
difference:

\no (a) {\bf Star formation}.
The near-supersonic cold streams provide a new efficient mechanism for
early star
formation. This is in contrast to the gradual infall of cooling shock-heated
gas assumed in most SAMs,
which starts from near rest, accretes smoothly into the disc, and joins the
quiescent mode of star formation there.
We find that the expected cold-gas supply is significantly more efficient
than assumed in most current models even in small haloes
(Cattaneo, Neistein, Birnboim \& Dekel, in preparation).

\no (b) {\bf Heating inside out}.
The concept of an expanding ``cooling radius" used in current SAMS is
limited to massive haloes where a virial shock exists.
Otherwise, it is the shock causing the heating which is propagating 
from the inside out.

\no (c) {\bf Shutdown of star formation}.
The combination of shock heating and AGN feedback provides a mechanism
for shutting off disc growth and star formation above a threshold halo
mass. 

\no (d) {\bf Cold streams}.
Cold streams that penetrate through the hot media continue to make discs 
and produce stars in haloes above $\msh$. This happens
mostly at $z \geq 2$, and preferentially in less grouped galaxies,
allowing big blue galaxies mostly at high $z$ and some at
low-density environments, and enforcing
a sharp shutdown of star formation at late times and especially
in clustered galaxies.

A practical schematic recipe for the critical halo mass
below which cold streams prevail and above which one may
assume a shutdown of gas supply and star formation is:  
\be
\mc = \cases{&$\!\!\!\!\!\msh ,\quad  z<\zc $\cr
         &$\!\!\!\!\!\msh \left(\frac{\msh}{f\mps(z)}\right) ,\quad z>\zc $ }
\ee
where the critical redshift $\zc$ is defined by $f\mps(\zc)=\msh$,
the clustering scale $\mps(z)$ is given by \equ{mstar}, 
and $f$ is a numerical factor of order a few.
Our best estimates for the parameters are
$\log \ms \simeq 11.8$ (but possibly another value in the
range 11.3-12.3) and $f \simeq 3$ (possible range: 1-10).
Using the approximation $\log \mps \simeq 13.1 - 1.3 z$ 
($z \leq 2$), we obtain $\zc \simeq 1.4$ for $f=3$.
This recipe should allow big blue systems at $z \geq 2$,
eliminate big blue systems and make big red galaxies at $z \leq 1$,
and generate a bi-modality near $\msh$. 
This scheme can be refined to allow for a smooth transition
above the critical scale by applying the shut-down to a varying
fraction of the gas and by breaking the streams into clumps which
will generate high peaks of starbursts. 

In addition, one may wish to have an effective minimum requirement for the
central black-hole mass 
in order to ensure enough feedback energy for maintaining the gas hot.
This may emphasize the bi-modality gap in color and bulge-to-disc ratio.
However, the proposed shut-down by halo mass may be enough for ensuring 
sufficient bulge mass and black-hole mass. 

The SAMs should be re-engineered to incorporate these processes
and thus help working out the detailed implications of the proposed scenario.
Preliminary attempts to do that, using two different SAMs, 
indicate that the incorporation of the new proposed processes
outlined above indeed leads to significantly better fits with the
observed bi-modality features along the lines proposed in \se{bimo}
(C05).

%---------------------
\subsection{Open issues}

In parallel, the physics of the involved ingredients should be
properly worked out in more detail, starting with the following
two hypotheses that were laid out in \se{bimo}.

%SF
\no (a) {\bf Fate of cold streams}.
A detailed investigation is required of the way by which
the cold streams evolve and
eventually merge with the central disc, the associated star formation,
and the resulting feedback process. While progress can be made using toy 
models and simplified simulations, a proper analysis  
will require simulations of higher resolution than are currently 
available. 
In particular,
whether or not the predicted star bursts could be associated with the big dusty
sources indicating massive star formation at high 
redshifts, such as the SCUBA sources \citep{chapman03_scuba},
remains to be determined once the theory is worked out
and the observed characteristics of these sources are clarified.

%AGN
\no (b) {\bf AGN feedback}.
The physics of AGN feedback is another unknown.
One wishes to understand 
how the available energy originating near the central black hole
is transferred to the hot gas spread over the halo.
The physics of how thermal conductivity may heat the gas
is also to be investigated. 
The increased efficiency of these feedback mechanisms in the presence of a hot
medium as opposed to their effect on cold flows and clumps are
to be quantified.

Parallel attempts to work out the details of the physical input
and to incorporate it in quantitative models of galaxy formation
will lead to progress in our understanding of the galaxy bi-modality
and the associated features.

%%%%%%%%%%%%%%
\section*{Acknowledgments}
We thank our collaborators A. Cattaneo, S.M. Faber, A. Kravtsov, E. Neistein, 
J.R. Primack, P. Seleson, R. Somerville \& E. Zinger.
We thank R. Dave, N. Katz, D. Keres \& D. Weinberg for sharing with us 
the results of their simulations.
We acknowledge stimulating discussions with
J. Binney, D. Lin, G. Kauffmann, G. Mamon, J.P. Ostriker, \& D. Weinberg.
This research has been supported by ISF 213/02 and NASA ATP NAG5-8218.
AD acknowledges support from a Miller Visiting Professorship at UC Berkeley,
a Visiting Professorship at UC Santa Cruz, and a Blaise Pascal International
Chair by Ecole Normale Superiere at the Institut d'Astrophysique, Paris.

%%%%%%%%%%%%%%%%%%%%%%%%%%%%%%%%%%%%%%%%%%%%%%%%%%

%\addtolength{\baselineskip}{-0.05\baselineskip}

\bibliographystyle{mn2e}

\bibliography{dekel}

%%\begin{thebibliography}{}

%%\def\refe{\bibitem}

%%\refe[Bardeen \etal(1986)]{bardeen:86}
%%Bardeen J. M., Bond J. R., Kaiser N., Szalay A. S., 1986, ApJ,  304, 15

%%\end{thebibliography}

%%%%%%%%%%%%%%%%%%%%%%%%%%%%%%%%%%%%%%%%%%%%%%%%%
\appendix
\section{Useful Relations}
\label{sec:app}

We summarize here the cosmological relations used in the analysis of
\se{shock_cos}. 
This is rather basic material, based for example on
\citet*{lahav91,carroll92} and \citet{mo02}.
By specifying it here in a concise and convenient form,  
we hope to allow the reader to reproduce our results and use them
in future work.
Additional relations associated with the spherical top-hat collapse
model are brought in the appendix of BD03.

\subsection{Cosmology}

The basic parameters characterizing a flat cosmological model in the 
matter era are the current values of the mean mass density parameter $\omm$
and the Hubble constant $H_0$.
At the time associated with expansion factor $a=1/(1+z)$, 
the vacuum-energy density parameter is $\oml(a)=1-\omm(a)$ and
\be
\omm(a)=\frac{\omm\, a^{-3}} {\oml +\omm a^{-3}} \ .
\ee 
The Hubble constant is
\be
H(a)=H_0\, (\oml +\omm a^{-3})^{1/2},
\ee 
and the age of the universe is
\be
\tu(a)=\frac{2}{3} H(a)^{-1}\,
\frac{ sinh^{-1}(|1-\omm(a)|/\omm(a))^{1/2} } {(|1-\omm(a)|)^{1/2}} \ .
\label{eq:tu}
\ee
%(with the $sinh^{-1}$ changing to a $sin^{-1}$ for $\omm>1$).
The mean mass density is
\be
\rho_{u} \simeq 1.88 \times 10^{-29} \Omega h^2 a^{-3}  %1.878
     \simeq 2.76 \times 10^{-30} {\omm}_{0.3}\, h_{0.7}^2\, a^{-3} \ , %2.761
\label{eq:rhou}
\ee
where ${\omm}_{0.3} \equiv \omm/0.3$,
$h\equiv H_0/100\,km\,s^{-1}\,Mpc^{-1}$,
and $h_{0.7}\equiv h/0.7$.

\subsection{Virial relations}

The virial relations between halo mass, velocity and radius,
\be
\Vv^2 = \frac{G\Mv}{\Rv} ,
\quad \frac{\Mv}{\frac{4\pi}{3} \Rv^3} = \Delta \rho_u
\ee
become
\be
M_{11} \simeq 6.06\, V_{100}^3 A^{3/2} % 6.064
    \simeq 342\, R_1^3 A^{-3} , % 341.77
\label{eq:virial}
\ee
where $M_{11} \equiv \Mv/10^{11}\msun$,
$V_{100} \equiv \Vv/100\kms$,
$R_1 \equiv \Rv/1\,Mpc$, and
\be
A\equiv (\Delta_{200}\, {\omm}_{0.3}\, h_{0.7}^2)^{-1/3}\, a .
\ee
An approximation for $\Delta(a)$ in a flat universe (Bryan \& Norman 1998) is:
\be
\Delta(a) \simeq (18\pi^2 -82\oml(a)-39\oml(a)^2)/\omm(a) \ .
\label{eq:Delta}
\ee

The virial temperature can be defined by
\be
\frac{k\Tv}{m}=\half \Vv ^2 \ .
\label{eq:Tv}
\ee
For an isotropic, isothermal sphere, this equals $\sigma^2$, where $\sigma$
is the one-dimensional velocity dispersion and the internal energy 
per unit mass is $e=(3/2) \sigma^2$.
Thus
\be
V_{100}^2 \simeq 2.79\, T_6 \, \quad   % 2.7866
M_{11} \simeq 28.2\, T_6^{3/2}\, A^{3/2}\ , % 28.207
\label{eq:tvir}
\ee
where $T_6 \equiv \Tv/10^6$K.

%----------------------------
\subsection{Press Schechter}

Linear fluctuation growth is given by 
\citep{lahav91,carroll92,mo02}
%(Lahav et al 1991, CPT92, MW02):
\be
D(a)=\frac{g(a)}{g(1)} \, a ,
\ee
where
\begin{eqnarray}
g(a) 
&\simeq& \frac{5}{2}\omm(a)\\
&\times&
\left[\omm(a)^{4/7}-\oml(a) +\frac{(1+\omm(a)/2)}{(1+\oml(a)/70)} \right]^{-1} .
\nonumber
\end{eqnarray}

The CDM power spectrum is approximated by
\citep{bbks86}:
\be
P(k) \propto k\, T^2(k) ,
\ee
with
\begin{eqnarray}
T(k)\!\!\!\!&=&\!\!\!\!\frac{\ln(1+2.34\,q)}{2.34\, q} \\
    \!\!\!\!&\times&\!\!\!\![1 +3.89q +(16.1q)^2 +(5.46q)^3 +(6.71q)^4]^{-1/4} ,
  \nonumber
\end{eqnarray}
where
\be
q=k/(\omm h^2 Mpc^{-1}) .
\ee
It is normalized by $\sigma_8$ at $R=8\hmpc$, where
\be
\sigma^2(R)=\frac{1}{2\pi} \int_0^\infty dk\, k^2\, P(k)\, {\tilde{W}}^2(kR) ,
\ee
and with the Fourier transform of the top-hat window function
\be
\tilde{W}(x)=3(\sin x -x\cos x)/x^3 .
\ee

In the Press Schechter (PS) approximation, the characteristic halo mass
$\mps(a)$ is defined as the mass of the 1-$\sigma$ fluctuation,
\be
1=\nu(M,a) =\frac{\delta_c}{D(a)\, \sigma(M)} ,
\quad \delta_c\simeq 1.69,
\label{eq:mstar}
\ee
where $M$ and the comoving radius $R$ are related via the universal density
today: $M=\frac{4\pi}{3} \bar\rho_0\, R^3$.
The mass of 2-$\sigma$ fluctuations is obtained by setting $\nu(M,a)=2$, etc.
Based on the improved formalism of \citet{sheth02},
the fraction of total mass in haloes of masses exceeding $M$ is
\be
F(>M,a) \simeq 0.4 \left( 1 +\frac{0.4}{\nu^{0.4}} \right)
{\rm erfc}\left( \frac{0.85\,\nu}{\sqrt{2}} \right) \ .
\label{eq:sheth}
\ee
This fraction for 1-$\sigma$, 2-$\sigma$, and 3-$\sigma$
fluctuations is 22\%, 4.7\%, and 0.54\% respectively.

\Fig{coolzm} shows the PS mass $\mps$ as a function of redshift.
For the standard $\Lambda$CDM with $\sigma_8=0.9$
its value at $z=0$ is $M_{*0}=1.36\times 10^{13}\msun$.
One can see that an excellent practical fit in the range $0\leq z \leq 2$
is provided by a power law in this semi-log plot:
$\log \mps \approx 13.134 - 1.3 z$.
At larger redshifts this gradually becomes an underestimate.
Trying to provide crude power-law approximations,
we find that $\mps \prop a^{4.2} \prop t^{3.5}$
are crude approximations in the range $0 \leq z \leq 1$,
and that $\mps \prop a^{5} \prop t^{4}$ are good to within a factor of 2
in the range $0 \leq z \leq 2$.
These power laws become overestimates at higher redshifts.

%%%%%%%%%%%%%%%%%%%%%%%%%%%%%%%%%%%%%%%%%%%%%%%%%%%

\label{lastpage}
\end{document}

%%%%%%%%%%%%%%%%%%%%%%%%%%%%5